\documentstyle[12pt,a4]{article}
\newcommand{\pr}{Phys.\ Rev.\ }
\newcommand{\prp}{Phys.\ Rep.\ }
\newcommand{\prep}{Phys.\ Rep.\ }

\newcommand{\np}{Nucl.\ Phys.\ }
\newcommand{\zp}{Z.\ Phys.\ }

\newcommand{\cpc}{Comput.\ Phys.\ Commun.\ }
\newcommand{\sjnp}{Sov.\ J.\ Nucl.\ Phys.\ }
\newcommand{\norm}[1]{{\protect\normalsize{#1}}}
\newcommand{\LAP}
{{\small E}\norm{N}{\large S}{\Large L}{\large A}\norm{P}{\small P}}

\newcommand{\vs}[1]{\rule[- #1 mm]{0mm}{#1 mm}}

\begin{document}
\renewcommand{\thefootnote}{\fnsymbol{footnote}}
\newpage
\pagestyle{empty}
\setcounter{page}{0}

%%%%%%%%%%%%%%%%%%%%%%%%%%%%%%%%%%
%%%%%%%%%%%  ENTETE ENSLAPP  %%%%%%%%%%%%
%%%%%%%%%%%%%%%%%%%%%%%%%%%%%%%%%%

%\input BoxedEPS
%\SetOzTeXEPSFSpecial
%\HideDisplacementBoxes
\null
\begin{minipage}{4.9cm}
\begin{center}
{\bf
G{\sc\bf roupe} d'A{\sc\bf nnecy}\\
\ \\
Laboratoire d'Annecy-le-Vieux de Physique des Particules}
\end{center}
\end{minipage}
\hfill
%\BoxedEPSF{enslapp.epsf}
\hfill
\begin{minipage}{4.2cm}
\begin{center}
{\bf
G{\sc\bf roupe} de L{\sc\bf yon}\\
\ \\
Ecole Normale Sup\'erieure de Lyon}
\end{center}
\end{minipage}

\begin{center}
\rule{14cm}{.42mm}
\end{center}
%%%%%%%%%%%%%%%%%%%%%%%%%%%%%%%%%%
%%%%%%%%%%%  ENTETE ENSLAPP FIN %%%%%%%%%%
%%%%%%%%%%%%%%%%%%%%%%%%%%%%%%%%%%

\begin{center}
{\Large {\bf CASCADE PARTICLES, NUCLEAR EVAPORATION,}}\\[0,3cm]
{\Large {\bf AND RESIDUAL NUCLEI IN HIGH ENERGY}}\\[0,3cm]
{\Large {\bf HADRON-NUCLEUS INTERACTIONS}}

\vs{5}

{\bf A. Ferrari, P.R. Sala}\\
{\em INFN, Sezione di Milano, Via Celoria 16, I-20133 Milano, Italy}

\vs{0,5}

{\bf J. Ranft}\\
{\em Laboratoire de Physique Th\'eorique \LAP\footnote{URA 14-36
du CNRS, associ\'ee \`a l'Ecole Normale Sup\'erieure de Lyon et \`a
l'Universit\'e de Savoie.} \\Groupe d'Annecy: LAPP, Chemin de Bellevue,
BP
110, \\F-74941
Annecy-le-Vieux Cedex, France.}

\vs{0,5}

{\em and}

\vs{0,5}

{\bf S. Roesler}\\
{\em Universit\"at Siegen, Fachbereich Physik, D-57068 Siegen, Germany.}

\end{center}
\vs{5}

\centerline{ {\bf Abstract}}

\indent

Based on a Monte Carlo realization of the Dual Parton Model we study the
production of target associated particles and of nuclear
fragments in high energy hadron--nucleus interactions.
A formation zone intranuclear cascade of low energy
secondaries inside the target nucleus is discussed. We calculate
excitation energies of residual nuclei left after the intranuclear
cascade process and treat their further disintegration by
introducing models for the evaporation of protons, neutrons, and light
fragments, high energy fission, and
by applying a Fermi Break-up model to light nuclear fragments. The
results are compared to data on target associated
particle production. We furthermore calculate cross sections for the
production of nuclear fragments.

\vfill
\rightline{\LAP-A-551/95}
\rightline{Siegen SI 95-09}
\rightline{September 1995}

\newpage
\pagestyle{plain}
\renewcommand{\thefootnote}{\arabic{footnote}}
%
%-------------------- Introduction (Sect. 1) ---------------------------
%
\section {Introduction}
The Dual Parton Model (DPM)~\cite{Capella94a} and Monte Carlo (MC)
implementations of this model for hadron-hadron~\cite{Aurenche92a,Bopp94a},
hadron-nucleus, and nucleus-nucleus collisions~\cite{Moehring91,Ranft94c}
have been quite successful in describing many aspects of hadron production
in high energy collisions. So far however, MC models
for hadron-nucleus and nucleus-nucleus collisions based on the DPM did
mainly describe the high energy component of newly created hadrons,
not the many particles resulting from the nuclear disintegration following
the high energy collisions.

Models for nuclear evaporation and fragmentation and for high
energy fission are however usually included in hadron cascade
models such as {\footnotesize FLUKA}~\cite{Fasso94a,Fasso94b,Aarnio94}
used for detector simulation and for the evaluation of radiation damage
to high energy accelerator and detector components. In the older models,
the nuclear excitation energy, which is the starting point for
the calculation of the nuclear disintegration, was often
introduced only in a phenomenological way~\cite{Ranft72}
or it was calculated on the basis of valid but simple intranuclear cascade
models~\cite{Bertini63,Bertini69} which are not applicable in the
multi-GeV energy range of present experiments.
The intranuclear cascade models have been greatly improved since
then~\cite{Fasso94a,Fasso94b,Aarnio94,Yariv81} and their range of
validity can be extended to higher energies due to the introduction of
the formation zone concept~\cite{Stodolski75,Ranft89}.

Here, we use the {\footnotesize DTUNUC}~\cite{Moehring91} and
{\footnotesize DPMJET-II}~\cite{Ranft94c} MC implementations of the DPM
for high energy hadron-nucleus and nucleus-nucleus collisions.
These MC models contain a formation zone intranuclear
cascade which is responsible for knocking out cascade protons and
neutrons of the residual nucleus. The cascade protons have energies
which are typical for the so called grey prongs observed in emulsion
experiments. Therefore, we are able to calculate the
nuclear excitation energy of the residual nucleus. In a second step, this
excitation energy is the basis for nuclear evaporation,
and high energy fission reactions.
In the present paper these mechanisms are investigated in hadron-nucleus
collisions. In a forthcoming publication we intend to extend these
studies to peripheral high energy nucleus-nucleus collisions.

In Sec.2 we describe the formation zone intranuclear cascade
model~\cite{Ranft88a} and focus on the calculation of excitation energies.
In Sec.3 models for evaporation and fragmentation are presented.
In Sec.4 we compare computed cross sections and multiplicities of
grey and black prong production to experimental data.
Furthermore, cross sections for the production of residual nuclei
are discussed. In Sec.5 we summarize our results.
\noindent
%
%-------------------- Sect. 2 ------------------------------------------
%
\section{The calculation of excitation energies in the formation zone
         intranuclear cascade model}
\subsection{The two-component Dual Parton Model for hadron-nucleus
            and nucleus-nucleus collisions}
The two-component DPM and its MC realizations
have been discussed in detail
in~\cite{Capella94a,Aurenche92a,Bopp94a,Moehring91,Ranft94c,Engel95a}.
Therefore, we summarize only briefly the main steps leading to the
multiparticle state, which is the starting point for the intranuclear
cascade and evaporation models being described in this paper.

The MC model for hadron-nucleus and nucleus-nucleus
interactions starts from an impulse approximation for the
nucleons of the interacting nuclei. The spatial initial configuration,
i.e. the positions of the nucleons in space-time in the rest system of
the corresponding nucleus, is sampled from standard density distributions.
For energies above 3-5~GeV/nucleon
the collision proceeds via $\nu$ elementary interactions
between $\nu_p$ and $\nu_t$ nucleons from the projectile and target,
resp. The values $\nu, \nu_p,$ and $\nu_t$ are sampled according to
Glauber's multiple scattering formalism using the MC algorithm
of~\cite{Shmakov89}. The particle production
is well described by the two-component DPM which is applied as in
hadron-hadron interactions~\cite{Aurenche92a,Engel95a,Bopp94b}.
As a result a system of chains
connecting partons of the nucleons involved in the scattering process is
formed. The chains are hadronized applying the model
{\footnotesize JETSET}~\cite{Sjostrand86,Sjostrand87a}.
The hadrons may than cause
intranuclear cascade processes, which are treated by the formation zone
intranuclear cascade model~\cite{Ranft88a}, an extension of the
intranuclear cascade model~\cite{Bertini63,Bertini69}.
At energies below 3-5~GeV/nucleon the formation zone intranuclear cascade
model itself provides a reasonable description of inelastic nuclear
collisions.

In the following we summarize the main ideas of the formation zone
intranuclear cascade model for hadron-nucleus interactions.
Modifications which have to be introduced to describe nucleus-nucleus
collisions will be discussed in a forthcoming paper.
The physical picture explaining the absence of the intranuclear cascade
at high energies is the concept of the formation zone~\cite{Stodolski75}.
It has been introduced in analogy to the
Landau-Pomeranchuk~\cite{Landau53} effect, which explains
the observation that electrons passing through material become more
penetrating at high energies. For the formation zone of an electron
with 4-momentum $p$ and energy $E$ upon
radiation of a photon with 4-momentum $k$ one obtains
\begin{equation}
\label{tauelec}
\tau=\frac{E}{k\cdot p}=\frac{E}{m}\frac{1}{\omega_e},
\end{equation}
where $\omega_e$ is the frequency of the photon in the rest frame of the
electron and $E/m$ is the time dilatation factor from the electron rest
frame to the laboratory.
Within the quark model, the states being
formed in the primary nucleon-nucleon interaction can be understood as
consisting of valence quarks only, i.e without the full system of sea
quarks, antiquarks, and gluons and have therefore a reduced probability
for hadronic interactions inside the nucleus~\cite{Ranft88a}.
The formation zone concept can be translated to hadron production as
follows~\cite{Ranft89}.
Denoting the 4-momenta of the projectile hadron $p_p$ and the secondary
hadron $p_s$ in the laboratory frame with
\begin{equation}
p_p=(E_p,0,0,\sqrt{E_p^2-m_p^2}), \qquad
p_s=(E_s,\vec{p}_{s\perp},\sqrt{E_s^2-m_s^2-\vec{p}_{s\perp}^2})
\end{equation}
and replacing in Eq.~(\ref{tauelec}) the electron momentum by $p_p$ and
the photon momentum by $p_s$,
%Denoting the 4-momentum of the projectile hadron with
%$p_p=(E_p,0,0,\sqrt{E_p^2-m_p^2})$, that of the secondary hadron with
%$p_s=(E_s,\vec{p}_{s\perp},\sqrt{E_s^2-m_s^2-\vec{p}_{s\perp}^2})$, and
%using Eq.~(\ref{tauelec})
%the hadron formation zone in the laboratory frame can be written as
the hadron formation zone reads for $E_p\gg m_p$
\begin{equation}
\tau_{\mbox{\scriptsize Lab}}=\frac{2E_s}{(m_px)^2+m_s^2+p_{s\perp}^2},
\qquad x=\frac{E_s}{E_p}.
\end{equation}
Since for most of the produced secondaries the term $(m_px)^2$ can be
neglected one can approximate
\begin{equation}
\tau_{\mbox{\scriptsize Lab}}\approx \gamma_s\tau_s, \qquad
\gamma_s=\frac{E_s}{m_s}.
\end{equation}
In the rest system of the secondary hadron $s$,
we define an average formation time $\tau_s$ needed to create a complete
hadronic state~\cite{Ranft88a,Ranft89}
\begin{equation}
\label{deffortim}
\tau_s=\tau_0\frac{m_s^2}{m_s^2+p_{s\perp}^2}.
\end{equation}
$\tau_0$ is a free parameter, which has to be determined by
comparing particle production within the model to experimental data.
Typical values are in the range from 1~fm/$c$ to 10~fm/$c$\footnote{In
Refs.~\cite{Moehring91,Ranft94c} $\tau_0$ was fixed to $\tau_0=5$~fm/$c$
whereas in Ref.~\cite{Ranft88a}  $\tau_0$=1-2~fm/$c$ was used.}.
{}From the comparisons discussed further below we find $\tau_0=2$~fm/$c$.
For each secondary we sample a formation time $\tau$ from an exponential
distribution~\cite{Bialas84} with an average value as given in
Eq.~(\ref{deffortim}).
As it was described in~\cite{Moehring91},
in our MC model we know the full
space-time history of the collision. In any particular Lorentz frame we
can follow the trajectories of the secondaries created in the
hadronization of the chains in space and time.
Due to relativistic time dilatation secondaries
with high energies in the nucleus rest system are mostly formed outside
the nucleus and are therefore not able to initiate intranuclear cascade
processes. On the other hand, the lower the energy of the secondary
hadronic system the higher is the probability to form a hadron
inside the nucleus. These hadrons may therefore reinteract with
spectator nucleons taking into account the nuclear geometry.
In the model, inelastic
secondary interactions of hadrons with energies below 9~GeV are
described with the code {\footnotesize HADRIN}~\cite{Haenssgen86}.
In general, the intranuclear cascade would start with resonances
resulting from the chain hadronization procedure, but we apply the
following way. Since the interaction cross sections of resonances
needed within {\footnotesize HADRIN} are less well known, we firstly
treat their decay
and sample the intranuclear cascade starting from the stable particles.
We assume that the effect of this approximation can mostly be
compensated by using an effective $\tau_0$ value.
%We assume that results obtained this way together with a certain
%$\tau_0$-value do not significantly differ from those which would be
%obtained starting the cascade from resonances together with different
%$\tau_0$-value.
Reinteractions within
the colliding nucleus beyond 9~GeV are very rare and therefore neglected
in the present approach. Pauli's principle is taken into account
as described in~\cite{Moehring91}.
%by
%checking the actual number of nucleons with momenta below the actual
%Fermi-threshold. Note, that cascade interactions may add or remove
%nucleons from the Fermi sea. If the resulting number of nucleons
%with energies below the Fermi-energy
%exceeds the number of nucleons in the initial state nucleus,
%the secondary interaction is forbidden.
For the secondaries produced in intranuclear cascade processes
we apply the same formalism as described
above and therefore obtain a formation zone intranuclear cascade in all
generations of secondaries.
\subsection{The calculation of nuclear excitation energies}
The treatment of nuclear effects within the MC model has
already been discussed in~\cite{Moehring91}. Since they are essential in
calculating excitation energies of nuclei left after primary
interactions and intranuclear cascade processes we summarize the
basic ideas.
Fermi momenta for nucleons as well as a simplified treatment of the
nuclear potential are applied to control the
generation of low-energy particles. Nucleon momenta are sampled from
zero-temperature Fermi distributions
\begin{equation}
\frac{dN^{\mbox{\scriptsize n,p}}}{dp}=N^{\mbox{\scriptsize n,p}}
\frac{3p^2}{(p_{\mbox{\scriptsize F}}^{\mbox{\scriptsize n,p}})^3}.
\end{equation}
Here and in the following the indices ``n'' and ``p'' denote neutrons
and protons, resp.
The maximum allowed Fermi momenta of neutrons and protons are
\begin{equation}
\label{maxFermi}
p_{\mbox{\scriptsize F}}^{\mbox{\scriptsize n,p}}=
\left[\left(\frac{N^{\mbox{\scriptsize n,p}}}{V_A}\right)
\frac{3h^3}{8\pi}\right]^{\frac{1}{3}}
\end{equation}
%\simeq0.4(\rho_A^{\mbox{\scriptsize n,p}})^{\frac{1}{3}} \mbox{GeV/}c,
with $V_A$ being the volume of the corresponding nucleus with an
approximate nuclear radius $R_A=r_0A^{1/3}, r_0=1.29$~fm.
%%r_0\approx 1$~fm.
%$\rho_A^{\mbox{\scriptsize n,p}}$ is the average density of protons
%or
%%and neutrons, resp.,
%neutrons
%within the nucleus.
Modifications of the actual nucleon momentum distribution, as they would
arise, for instance, taking the reduced density and momenta in the nuclear
skin into consideration,
%as discussed in~\cite{Bertini63}
effectively result in a
reduction of the Fermi momenta as compared to those sampled from
Eq.~(\ref{maxFermi}). This effect can be estimated by a correction
factor $\alpha_{\mbox{\scriptsize mod}}^{\mbox{\scriptsize F}}$ which
modifies the Fermi-momenta.
Results presented in this paper have been obtained with
$\alpha_{\mbox{\scriptsize mod}}^{\mbox{\scriptsize F}}$=0.75.
The depth of the nuclear potential is assumed to be the Fermi energy and the
binding energy for outer shell nucleons
\begin{equation}
\label{nucpot}
V^{\mbox{\scriptsize n,p}}=
\frac{(p_{\mbox{\scriptsize F}}^{\mbox{\scriptsize n,p}})^2}
{2m_{\mbox{\scriptsize n,p}}}+
E^{\mbox{\scriptsize n,p}}_{\mbox{\scriptsize bind}}.
\end{equation}
To extend the applicability of the model to the energy region well below
1~GeV an approximate treatment of the Coulomb-potential is provided.
The Coulomb-barrier modifying the nuclear potential is calculated from
\begin{equation}
\label{coulpot}
V_{\mbox{\scriptsize C}} = \frac{e^2}{4\pi\epsilon_0 r_0}
\frac{Z_1Z_2}{(A_1^{1/3}+A_2^{1/3})}
\end{equation}
with the mass numbers $A_1, A_2$ and charges $Z_1, Z_2$ of the
colliding nuclei, i.e. with $A_1=|Z_1|=1$ for charged hadrons entering
or leaving the target nucleus. $e$ denotes the elementary charge and
$r_0=1.29$~fm.

The excitation energy $U$ of the residual nucleus with mass number
$A_{\mbox{\scriptsize res}}$
and charge $Z_{\mbox{\scriptsize res}}$, i.e. the energy above the
ground state mass $E_{0,\mbox{\scriptsize res}}$, is given as
\begin{eqnarray}
&U&=E_{\mbox{\scriptsize res}}-E_{0,\mbox{\scriptsize res}},\nonumber\\
&E_{0,\mbox{\scriptsize res}}&=
Z_{\mbox{\scriptsize res}}m_{\mbox{\scriptsize p}}+
(A_{\mbox{\scriptsize res}}-Z_{\mbox{\scriptsize res}})m_{\mbox{\scriptsize n}}
-E_{\mbox{\scriptsize bind}}
(A_{\mbox{\scriptsize res}},Z_{\mbox{\scriptsize res}}).
%,\qquad m_{\mbox{\scriptsize n}}=0.9315\mbox{~GeV}
\end{eqnarray}
We calculate the binding energy
$E_{\mbox{\scriptsize bind}}
(A_{\mbox{\scriptsize res}},Z_{\mbox{\scriptsize res}})$
using the experimentally determined excess masses of all known (measured)
nuclides and using mass formulae for
nuclides far from the stable region, where no measurements are available.
The excitation energy is obtained within our model from an explicit
consideration of the effects of
the nuclear potential (Eq.~(\ref{nucpot})) and the
Coulomb energy (Eq.~(\ref{coulpot})), i.e. from corrections
which are applied to the
4-momenta of the final state hadrons leaving the spectator nucleus.
We modify the energies of these hadrons by the potential barrier and
rescale the 3-momenta correspondingly. It is assumed
that these corrections have to be applied to nucleons wounded in
primary and secondary interactions and to those hadrons only, which are
formed inside the spectator nucleus corresponding to the sampled
formation time. Among these particles we find apart from the nucleons a
small fraction of other baryons, which are assumed to move in a nucleon
potential and mesons to which we apply an effective meson potential
of 0.002~GeV. Due to energy-momentum conservation these
corrections lead to a recoil momentum and, therefore,
to an excitation of the residual nucleus. In addition, there is
a further contribution to the recoil momentum of the residual nucleus
arising from potential corrections applied to the momentum of the
projectile hadron entering the nuclear potential and from cascade
nucleons with kinetic energies below the nuclear potential which are
therefore not able to escape the spectator nucleus.

In Fig.\ref{pAex1}a we show the dependence of the average excitation
energies of the target residual nuclei on the momenta of the projectile
in the laboratory. The decrease of the excitation energy for momenta
below about 50~GeV/$c$ is mainly due to the breakdown of the Glauber
cascade\footnote{Note, that the Glauber cascade as obtained with
Glauber's formalism is biased by sampling the actual chain
systems~\cite{Moehring91}. In order to ensure that the chain masses
$M^2_{\mbox{\scriptsize chain}}=sx_{p}x_{t}$ exceed the masses of the
lowest-mass
hadronic states with the corresponding quantum numbers lower $x$-cuts
are imposed for all parton systems. Therefore, at low energies these
$x$-cuts may reduce the number of sea quark containing chains.}
as it can be clearly seen for Au and Pb targets. This is not the
case for light nuclei where even at high energies only up to 2-3 target
nucleons are involved on average in the primary interaction. This
threshold behaviour will be discussed with respect to experimental
information on grey and heavy particles in more detail further below. At high
energies the average excitation energies are almost independent on the
projectile momentum as one would expect from limiting
fragmentation~\cite{Benecke69}.
The average excitation energies per nucleon of the residual target
nucleus are given in Fig.\ref{pAex1}b. In difference to
Fig.\ref{pAex1}a the $p_{\mbox{\scriptsize Lab}}$-dependence is
similar for all target
nuclei apart from a constant shift towards higher excitation energies
per nucleon for light nuclei. This is due to a smaller ratio of wounded
nucleons to all target nucleons for heavy nuclei as compared to light
nuclei.

In Fig.\ref{pAex2} we show the average excitation energy of the
residual target nucleus depending on the mass number of the target for
proton-nucleus interactions at 300~GeV/$c$. The
different symbols correspond to several numbers of nucleons lost by the
target in primary and secondary interactions, i.e. $\Delta A$ is defined
by $\Delta A=A-A_{\mbox{\scriptsize res}}$. The excitation energy is
strongly correlated
to the number of removed nucleons. The more nucleons one removes from
the target, the more energy is deposited into the spectator nucleus.
For a fixed number $\Delta A$ the excitation energy is increasing with
the mass number of the target.
The reason for this is, that in heavy targets we need more cascading
to remove a given number of nucleons than in light targets.
As an example, the distribution of
excitation energies and of excitation energies per nucleon of the target
prefragment are shown for proton-gold interactions at 300~GeV/$c$
in Fig.\ref{pau300ex}a,b. In addition to the distributions obtained
taking all prefragments into account
(labelled ``all $A_{\mbox{\scriptsize res}}$'') we
give distributions which correspond to several mass ranges, in
particular to several lower cuts in the prefragment mass. Again, the
more nucleons are involved in the primary interaction and the
intranuclear cascade the higher are the mean excitation energies and,
therefore, the broader are the excitation energy distributions.
%
%-------------------- Sect. 3 ------------------------------------------
%
\section{\label{evamodel}Evaporation/Fragmentation}
At the end of the intranuclear cascade the residual nucleus is supposed to
be left in an equilibrium state, in which the excitation energy $U$ is shared
by a large number of nucleons. Such an equilibrated compound nucleus is
supposed to be characterized by its mass, charge, and excitation energy
with no further memory of the steps which led to its formation.
The excitation energy can be higher than the separation energy,
thus nucleons and light fragments ($\alpha$,d,$^3$H, $^3$He) can
still be emitted: they constitute the low-energy (and most
abundant) part of the emitted particles in the rest system of the
residual nucleus, having an average energy of few MeV.
The emission process can be well described as an evaporation from a hot
system. The treatment starts from the formula of
Weisskopf~\cite{Weisskopf37}, that is an application of the detailed balance
principle.
The evaporation probability for a particle of type $j$ , mass $m_j$,
spin $S_j \cdot \hbar$, and kinetic energy $E$ is given by
\begin{equation}
\label{eq:Weisskopf}
P_j(E)dE=\frac{(2S_j+1)m_j}{\pi^2\hbar^3}\sigma_{\mbox{\scriptsize inv}}
\frac{\rho_f(U_f)}{\rho_i(U_i)}EdE
\end{equation}
where $\rho$'s are the nuclear level densities ($\rho_f(U_f)$ for the
final nucleus, $\rho_i(U_i)$ for the initial one), $U_i \equiv U$ is the
excitation energy of the evaporating nucleus, $U_f=U-E-Q_j$ that of
the final one, $Q_j$ is the reaction $Q$ for emitting a particle of
type $j$ from the original compound nucleus, and
$\sigma_{\mbox{\scriptsize inv}}$ is the cross section for the inverse
process.

Eq.~(\ref{eq:Weisskopf}) must be implemented with a suitable form for the
nuclear level density and the inverse cross sections. Many recipes have
been suggested for both. In the original work of
Dostrovsky~\cite{Dostrovsky59},
$\rho(U)\approx C\exp{(2\sqrt{aU})}$, with $a=A/8$
has been used for the level density dependence on the excitation
energy~$U$.
This has led to a simple form for the evaporation probability:
\begin{equation}
\label{eq:Dos}
P_j(E)dE=\frac{(2S_j+1)m_j}{\pi^2\hbar^3}\sigma_{\mbox{\scriptsize inv}}
\frac{e^{2\sqrt{a(U-E-Q_j)}}}
{e^{2\sqrt{aU}}}EdE.
\end{equation}
In the same work, the
inverse cross sections have been parametrized in a very simple way, so
that expression (\ref{eq:Dos}) can be analytically integrated and used for
MC sampling. The same formulation is used in this work with,
however, a different choice of $a$ as it will be discussed later.

The total width for neutron emission can be found by integrating
Eq.~(\ref{eq:Weisskopf}) between zero and the maximum possible ejectile energy
$(U-Q_j)$
\begin{equation}
\label{eq:gammaj}
\Gamma_j=
\frac{(2S_j+1)m_j}{\pi^2\hbar^2}
\int_0^{(U-Q_j)}{\sigma_{\mbox{\scriptsize inv}}(E)\frac{\rho_f}{\rho_i}EdE}.
\end{equation}
The same applies to charged particles, where the integration actually
goes from some effective Coulomb barrier where
$\sigma_{\mbox{\scriptsize inv}}$ drops to zero, up to the maximum energy.

The evaporative process is in competition with another equilibrium
process, that is fission~\cite{Vandenbosh73}.
For the fission probability,
a statistical method can be used~\cite{Weisskopf37,Bohr39}:
obtaining for the total fission width
\begin{equation}
\label{eq:gammaf}
\Gamma_{\mbox{\scriptsize F}}=
\frac{1}{2\pi}\frac{1}{\rho_i(U)}\int_0^{(U-B_{\mbox{\scriptsize F}})}
{\rho_{\mbox{\scriptsize F}}(U-B_{\mbox{\scriptsize F}}-E)dE}
\end{equation}
where $B_{\mbox{\scriptsize F}}$ is the fission barrier, and
$\rho_{\mbox{\scriptsize F}}(U_{\mbox{\scriptsize F}})\approx
C\exp{(2\sqrt{a_{\mbox{\scriptsize F}} U_{\mbox{\scriptsize F}}})}$,
the level density of the
fissioning nucleus at the saddle point, where the excitation energy
$U_{\mbox{\scriptsize F}}$ is given by the initial one minus the fission
barrier.

We follow the prescriptions of Atchison~\cite{Atchison80}
to calculate the quantities entering Eq.~(\ref{eq:gammaf}),
except, again, for the level density parameter $a_{\mbox{\scriptsize F}}$.

In both $\rho_{\mbox{\scriptsize F}}(U)$ and $\rho(U)$ we use the
so-called backshifted level
density, using $U-\Delta$ rather than $U$, where $\Delta$ is the
pairing energy. Moreover, $\tilde a = a/A$, and
$\tilde a_{\mbox{\scriptsize F}}=a_{\mbox{\scriptsize F}}/A$
are found to be all but constant parameters: they possess a dependence
on $A$ and $Z$, due to shell and deformation effects,
and a dependence on excitation energy. Both effects
have been experimentally observed, and have been
subject of many phenomenological and theoretical investigations
(see~\cite{Gilbert65,Ignatyuk75a,Ignatyuk75b,Mashnik93,Iljinov92,Shlomo92}).
Here the $N$ and $Z$ dependence of Ref.~\cite{Gilbert65} is used, and
complemented with the energy dependence prescription of
Ignatyuk~\cite{Ignatyuk75a,Ignatyuk75b}
\begin{eqnarray}
\label{eq:camign}
a &=& A \cdot \left[ \bar{a} \cdot f(U) + \tilde{a} \cdot \left( 1 - f(U)
\right) \right] \nonumber \\
\bar{a} &=& a_0 + 9.17 \times 10^{-3} \cdot \left[
S_{\mbox{\scriptsize Z}}(Z) + S_{\mbox{\scriptsize N}}(N) \right]
\\
\tilde{a} &=& 0.154 - 6.3\times 10^{-5}\cdot A \nonumber \\
f(U) &=& \frac{ 1 - e^{-0.054\cdot (U-\Delta)}}{0.054 \cdot
(U-\Delta) } \nonumber
\end{eqnarray}
where according to~\cite{Gilbert65}, $a_0$ is given by 0.142~MeV$^{-1}$ and
0.12~MeV$^{-1}$ for undeformed and deformed nuclei respectively, and
$S_{\mbox{\scriptsize Z}}(Z)$ and $S_{\mbox{\scriptsize N}}(N)$ are
the shell correction terms for protons and neutrons. The unit of energy
used throughout Eq.~(\ref{eq:camign}) is MeV.

The level density at the saddle point $\rho_{\mbox{\scriptsize F}}$
is different from that of
the nucleus in its ground state. From
comparison to experimental data, it turns out that
$a_{\mbox{\scriptsize F}}$ is greater than
the $a$ used for evaporation of about 10\% at low excitation energies,
and the two $a$'s become equal at large excitation energies.
We use $a_{\mbox{\scriptsize F}} \approx
1.08 a$, with a smooth $A$ dependence. After fission occurs, the two
fragments are treated like independent residual nuclei with their own
excitation and can possibly emit further particles.

For light nuclei, the statistical assumptions and the
sequential emission scheme underlying the classical evaporation models
become less and less applicable, because:
\begin{itemize}
\item Already moderate excitation energies can represent a substantial
fraction of the (total) binding energy of such nuclei.
\item The level structure of such nuclei is usually highly specific and
anyway level spacings can be comparable with the excitation energy.
\item The ``evaporation'' of light fragments other than p or n becomes
meaningless, since the mass of the ``evaporated'' fragment can be
comparable or even larger than the mass of the residual nucleus.
\end{itemize}
Therefore other deexcitation mechanisms are more suitable for these
light residual nuclei. The one adopted for this calculations
is the so called Fermi Break-up model~\cite{Fermi50,Epherre67},
 where the excited nucleus is supposed to
disassemble just in one step into two or more fragments, with branching
given by plain phase space considerations. In particular, the
probability for disassembling a nucleus of $N$ neutrons, $Z$ protons,
and $U$ excitation energy (total mass $M^*=U+M_{A,Z}$)
into $n$ fragments ($n\ge 2$) of the same
total charge and baryon number, is given by:
\begin{equation}
W=\frac{g}{G}\left[\frac{V_{\mbox{\scriptsize br}}}
{(2\pi\hbar)^3}\right]^{n-1}
\left(\frac{1}{M^*} \prod_{i=1}^n m_i \right)^{3/2}
\frac{(2\pi)^{3(n-1)/2}} {\Gamma(\frac{3}{2}(n-1))}
E_{\mbox{\scriptsize kin}}^{3n/2-5/2}
\end{equation}
where the spin factor $g$, and the permutation factor $G$ are given by
($n_j$ is the number of identical particles of $j$th kind)
\begin{equation}
g=\prod_{i=1}^n(2S_i+1),\qquad G=\prod_{j=1}^{k} n_j!
\end{equation}
and $E_{\mbox{\scriptsize kin}}$ is the total kinetic energy of all
fragments at the moment
of break-up. $V_{\mbox{\scriptsize br}}$ is a volume of the order of the
initial residual nucleus volume.
Therefore, the final state are conveniently selected by means of a
MC procedure, by evaluating such an expression for all possible
combinations of fragments energetically allowed
and making a random selection.
We considered all combinations formed by  up to six fragments,
unless the residual ``nucleus'' is
composed by $A$ like particles (p or n), in which case it is disintegrated
into $A$ fragments according to phase space.
All particle stable states with $A\le 16$ have been included, plus the particle
unstable levels with sizeable $\gamma$ decay branching ratios. Also a
few known particle unstable isotopes, like $^8$Be, have been included
and, if produced, are let to decay according to the experimental
branching.
Once the final state configuration has been selected, the kinematical
quantities of each fragment are chosen according to $n$-body phase
space distribution. Such a selection must be performed taking care to
subtract from the available energy the Coulomb repulsion of all charged
particles: the Coulomb energy is then  added back to the charged
particles alone, to simulate properly the effect of the Coulomb
repulsion. In practice $E_{\mbox{\scriptsize kin}}$ at disassembling
will be given by:
\begin{equation}
E_{\mbox{\scriptsize kin}}=U-\left(\sum_{i=1}^n m_i-M_{A,Z}\right)-
B_{\mbox{\scriptsize Coul}}
\end{equation}
where it must be recalled that the emitted fragments can be in an
excited state. The total Coulomb barrier $B_{\mbox{\scriptsize Coul}}$
of the selected
configuration is distributed to charged particles after disassembling,
in their own c.m. system.

According to the picture of the compound nucleus like an equilibrated
system determined only by its mass, charge and excitation energy, with no
memory of previous steps of the interaction, Fermi Break-up is activated
in the model every time the current compound nucleus has mass number
$A\le 17$, including possible light fission fragments.
The fragmentation of higher mass compound nuclei is not yet included
in the model. This process, although its cross section is quite
small, is important when considering the distribution of residual
nuclei, because it can produce isotopes very far both from the target
mass and from the fission product distribution.
%
%-------------------- Sect. 4 ------------------------------------------
%
\section{The production of grey and black particles
         and residual nuclei in high energy collisions}
\subsection{\label{slowprod} Grey and black particles and correlations}
The intranuclear cascade of low energy secondaries and the evaporation
of nucleons and light fragments mainly contribute to the production of
hadrons and light fragments with a velocity less than about $0.7c$ in the
rest frame of the target nucleus. They are frequently called ``target
associated'' or ``slow'' particles. Most of the experimental information
on slow particle production presently comes from experiments using
nuclear emulsions as targets. The emulsions usually consist of a
component of light nuclei (H,C,N,O) and a component of heavy nuclei
(Ag,Br). The appearance of slow particles in these experiments has
led to their subdivision into ``grey'' and ``black'' particles.
The exact definition usually differs slightly between different
experiments. As shown by many authors this subdivision has not only an
experimental meaning but also subdivides slow particle production into
a region which can be understood by intranuclear cascade processes (grey
particle production) and a region of black particle production based on
nuclear evaporation processes. If not explicitly stated, throughout this
paper we apply the following definitions:
grey particles are assumed to be singly charged particles with a
Lorentz-$\beta$ value between 0.23 and 0.7 and black particles
are singly and multiply charged particles with
$\beta<0.23$. This is in agreement with definitions usually assumed in
experiments~\cite{Stenlund82,ALMT75}. Furthermore, within our calculation we
use the emulsion-composition from~\cite{Stenlund82}, i.e. an emulsion
consisting of 28.8\% of light nuclei and of 71.4\% of heavy
nuclei.

In Tab.\ref{pemumulttab} we give the average multiplicities of grey, black, and
heavy (=grey+black) particles in proton-emulsion interactions as
obtained with our MC model {\footnotesize DTUNUC}~2.0 together with
experimental
results for different momenta of the projectile proton. In addition
this is shown in Fig.\ref{pemumult} together with the corresponding
shower particle ($\beta>0.7$) multiplicities. Whereas the average number of
shower particles is increasing throughout the whole energy range we
get an increasing multiplicity of grey and black particles up to
about 40~GeV/$c$ which turns into an almost constant behaviour for higher
energies.
Within our model, this constant behaviour is due to
limiting fragmentation in each hadron-nucleon interaction~\cite{Benecke69}
together with a constant formation zone intranuclear cascade and inelastic
hadron-nucleus cross sections depending only weakly on the projectile
energy.
%cascade after saturation imposed by the volume of the nucleus and by the
%distribution of nucleons within the nucleus has been reached
%and, therefore, to constant mean excitation energies of the residual nuclei
%(see Fig.\ref{pAex1}).
The position in energy of the
threshold region, i.e. the region of increasing heavy particle
multiplicities, is governed by the nuclear geometry and the nuclear
potential, which both are treated in a very rough manner, and by the
way the Glauber cascade is biased by lower cuts applied to
chain masses. Further shifts of this region or changes of the
slope in the threshold
region can be obtained by varying the distributions from which
the x-values of the sea-partons are
sampled\footnote{Within {\footnotesize DTUNUC} they
are sampled from a $1/x$-distribution~\cite{Moehring91}.}. The
experimental data are taken from a compilation of data by Fredriksson et
al.~\cite{Fredriksson87}, i.e. they were obtained in different
experiments. The definitions of ``grey'' and ``black'' may therefore
slightly vary between them. This fact may also partly account for the
fluctuations within the data for grey and black particle multiplicities
in the high energy region.
For momenta above  about 20~GeV/$c$ the model agrees
well with measured multiplicities, whereas at low energies our results
seem to depart from the experiments. However, the different definitions
used for ``grey'' and ``black'' cannot completely explain the big
differences within the experimental results at low energies and any
clear experimental information on the threshold behaviour is missing.
{}From this comparison it is therefore difficult to draw
conclusions about the applicability of our model of slow particle
production to energies below 10-20~GeV and to modifications within the
model which could be necessary to reproduce the observed threshold
behaviour.

In order to investigate how the two emulsion components
contribute to the energy dependence of the average
multiplicities discussed so far we show
in Tabs.\ref{pcnomulttab} and \ref{pagbrmulttab}
and, together with shower particle multiplicities, in Fig.\ref{pcnoagbrmult}
the mean grey, black, and heavy
particle multiplicities for the light and heavy component
separately in comparison to experimental data~\cite{Fredriksson87}.
In order to
illustrate the uncertainties within the experimental data we give for
some energies several multiplicity values, which were measured in
different experiments. As it is clearly shown in
Fig.\ref{pcnoagbrmult}a in interactions of protons with light nuclei
even at high energies only up to three target nucleons are interacting
with the projectile, i.e. we are dealing with a very limited Glauber
cascade and, therefore, do not observe the typical breakdown of the
cascade at low energies which would
manifest itself in decreasing grey and black particle multiplicities.
In agreement with the measurements we obtain constant
mean grey and black particle multiplicities in the whole energy range.
It seems, that the model overestimates the black particle multiplicity
which could be due to the absence of the treatment of the nuclear skin
in the nuclear potential, i.e. by underestimating the low part of the
excitation energy distribution.
However we must note, that the experiments were usually classifying the
target nuclei as belonging to one of the components by the number of
produced heavy particles, which implies further uncertainties. An
average value of one grey particle per interaction
agrees well with the experimental results.
The model reproduces the measured multiplicities of slow particles in
interactions of protons with nuclei of the heavy component
(Tab.\ref{pagbrmulttab}, Fig.\ref{pcnoagbrmult}b) down to a proton
momentum of about 20~GeV/$c$.

In Figs.\ref{hemung} and~\ref{hemunb} we present the grey and black
particle multiplicity distributions normalized to unity for proton-emulsion
interactions at 200~GeV (a) and $\Sigma^-$-emulsion interactions at
350~GeV (b) together with data~\cite{Stenlund82,Szarska93}.
As the comparisons show, our
model is able to reproduce the data on slow particle multiplicities very
well. The grey particle multiplicity distribution for proton projectiles
(Fig.\ref{hemung}a) slightly underestimates the measured distribution
at high multiplicities which, however, might be not very conclusive
since the uncertainties within the experimental data are rather big in
this region. The hump in the calculated black particle distribution
for proton projectiles at $N_b\approx 4$ (Fig.\ref{hemunb}a)
is due to the evaporation of charged particles from light
emulsion nuclei and seems to be less pronounced in the measured
distribution. This is not the
case in the $\Sigma^-$-emulsion data (Fig.\ref{hemunb}b), where, on the
other hand, the uncertainties are higher than within the proton-emulsion
data.

The reasonable description of slow particle multiplicity distributions
implies that the model should be able to reproduce measured correlations
between grey, black, and shower particle multiplicities. In
Fig.\ref{pemunbnscorr} we compare correlations between grey and shower
particle multiplicities (a,b) and between black and shower particle
multiplicities (c,d) and in Fig.\ref{pemungnbcorr}a,b
between grey and black particle multiplicities with data of the
Alma-Ata--Leningrad--Moscow--Tashkent Collab.~\cite{ALMT75} on
proton-emulsion interactions at 200~GeV and in Fig.\ref{pemungnbcorr}b
in addition to data of the KLM-Collab.~\cite{Dabrowska93a}.
In Ref.~\cite{ALMT75} the errorbars
are obviously only given for selected data points.
Apart from the correlation between grey and black particles, where we
obtain slightly more black particles for a fixed number of grey
particles than seen in the experiments, our
calculations are in good agreement with the data within their uncertainties.

A detailed experimental study of slow particle production in
interactions of protons, pions, and kaons with different target nuclei
at energies varying between 50~GeV and 150~GeV was presented
in~\cite{Braune82}. Here, grey particles are defined as charged
particles having a velocity between $0.3c$ and $0.7c$. In Tab.\ref{hAmulttab}
we compare our results on mean grey particle multiplicities to these data.
Again, the agreement is satisfactorily.

The dependence of the mean grey, heavy, and shower particle
multiplicities on the mass number of the target nucleus in
proton-nucleus collisions at 300~GeV/$c$ was subject to further
comparisons. The results are given in Tab.\ref{pAmulttab} and
Fig.\ref{pAmult} together with data taken from Ref.~\cite{Fredriksson87}.
Our model is reproducing the increase of the heavy
particle multiplicity with the target mass number. As mentioned above
the data point for C,N,O was obtained in emulsion
experiments in which the identities of the target nuclei were deduced from
the heavy particle multiplicities. Therefore this data point has to
be taken with care.

Finally, we compare grey particle multiplicity distributions in interactions
of protons and pions with different target nuclei at 200~GeV/$c$ with recent
data of the WA80-Collab.~\cite{Albrecht93}. In agreement with the
experiment grey particles are defined as singly charged particles with a
kinetic energy between 30~MeV and 400~MeV emitted in the target rapidity
region ($-1.7<\eta<1.3$). The result of the comparison is shown in
Fig.\ref{pAngwa80}. All calculated distributions are normalized to the
Glauber cross sections of the corresponding interactions. For the two
light nuclei (C,Al) our distributions are consistently broader than the
experimental distributions, whereas for heavier targets we agree well in
shape and absolute normalization with the data.
\subsection{Residual nuclei and high energy fission}
After evaporation most of the residual nuclei have lost up to one-third
of their nucleons depending on their mass $A_T$, on the kind and energy
of the projectile, and on the interaction characteristics (impact
parameter etc.). They may be considered as heavy fragments produced in a
spallation or a deep spallation process. In addition, the high energy
fission model and the Fermi Break-up model which were introduced in
Sect.\ref{evamodel} modify the mass spectrum of the nuclear
prefragments furthermore. In Fig.\ref{pagresA} we show the isobaric
mass yields of fragments in interactions of silver
nuclei with 11.5~GeV (a) and 300~GeV (b)
protons together with data~\cite{English74,Porile79}. Since
multifragmentation is not included in our MC model we get -- apart
from light fragments ($A\le 4$) which were evaporated from the residual
nucleus -- almost no fragments with masses below
$A_{\mbox{\scriptsize res}}\approx 40$. In the
spallation region ($50\le A_{\mbox{\scriptsize res}}\le 100$)
our calculation agrees within a factor of two with the
measured mass yields, which is satisfactorily in view of our simplified
approach and taking into account the fact that multifragmentation would
lead to a further decrease of the cross section. The rising yields of
fragments close to the target mass
($A_T-5\le A_{\mbox{\scriptsize res}}\le A_T$) are not
described within our model.
This is due to the fact that such processes like quasi-elastic
scattering are not treated within our model and it might be due to our
rough treatment of the nuclear potential, i.e. we probably underestimate
the low part of the excitation energy distribution by neglecting the
nuclear skin effects.
As experimental results on isotope production show, fragment
production cross sections remain about constant for projectile energies above
10~GeV (see Fig.\ref{pagresA} and~\cite{Kaufman76,Huefner85}). This
fact suggests that the regime of constant slow particle production may
already be
reached at an energy of about 10~GeV. In contrast, within our model the
threshold above which slow particle production does not change
significantly is at about 20--30~GeV (cf. discussion in Sect.\ref{slowprod}).
This fact explains the different shape of the calculated mass yields at
$E_{\mbox{\scriptsize Lab}}=11.5$~GeV (Fig.\ref{pagresA}a) as compared to
$E_{\mbox{\scriptsize Lab}}=300$~GeV (Fig.\ref{pagresA}b).
However, in order to draw further conclusions on the
threshold region which are based on fragment production cross sections,
it would be necessary to describe all aspects of the fragmentation
process (such as multifragmentation) which is beyond the scope of this
work.

In Fig.\ref{pauresA} we compare the charge yield obtained in
interactions of 10.6~GeV protons on $^{197}$Au nuclei (a) and the
isobaric mass yield obtained in interactions of 800~GeV protons on
$^{197}$Au nuclei (b) to data~\cite{Heinrich95,Sihver92}.
Since high energy fission
significantly modifies the fragment production cross sections we show
both, the mass yields of the residual nuclei after the evaporation-step
without high energy fission
(crosses) and mass yields obtained taking high energy fission into
account (diamonds). Within the limitations of our models we are able to
reproduce the measured yields very well, especially the yields at
$E_{\mbox{\scriptsize Lab}}=800$~GeV (Fig.\ref{pauresA}b) where we agree
with the data in
the mass range $60\le A_{\mbox{\scriptsize res}}\le 190$ within their
uncertainties. The calculated yields at
$A_{\mbox{\scriptsize res}}=2,3,4$ represent light fragments evaporated
from the prefragments.
Again, our models do not cover the multifragmentation
region and the mass region very close to the target mass.

In order to investigate isotope-production in more detail we compare
independent mass yield distributions from interactions of 800~GeV
protons with $^{197}$Au with data~\cite{Sihver92} in Fig.\ref{pAuZiso}.
There we plot the cross sections for the production of certain isotopes
with masses $A_{\mbox{\scriptsize res}}$ and charge $Z$ versus the
difference of their
charge and the most probable charge $Z_{\mbox{\scriptsize mp}}$ for
three intervals of
$A_{\mbox{\scriptsize res}}$. Corresponding to~\cite{Sihver92}
$Z_{\mbox{\scriptsize mp}}$ is defined as
\begin{equation}
Z_{\mbox{\scriptsize mp}}(A)=aA^2+bA+c
\end{equation}
with $a=-0.382\cdot 10^{-3}$, $b=0.483$, $c=0$ for $82\le
A_{\mbox{\scriptsize res}}\le 89$,
and $c=0.231$ for $122\le A_{\mbox{\scriptsize res}}\le 129$.
In the highest mass range
$166\le A_{\mbox{\scriptsize res}}\le 176$ $c$ had to be modified by
1.0 in order to compare
the shape of the distributions, i.e. $c=1.254$.
For each interval we calculate the independent yields for three
different $A_{\mbox{\scriptsize res}}$ values.
We are able to reproduce the measured charge distributions which
have the typical gaussian shape.

The average recoil momenta of the fragments in proton-$^{197}$Au
interactions at 800~GeV as a function of the mass loss
$\Delta A=A_T-A_{\mbox{\scriptsize res}}$ are shown together with data
from different
experiments~\cite{Morrissey89} in Fig.\ref{pAuprcl}. The momenta of the
fragments obtained with our MC model are in reasonable
agreement with the data.
%
% Finally, in Fig.\ref{palresA} we compare isobaric mass yields for
% interactions of 300~GeV protons on aluminum calculated applying the
% Fermi Break-up model for light fragments to the yields obtained without
% Fermi Break-up.

%
%-------------------- Summary (Sect. 5) --------------------------------
%
\section{Summary and conclusions}
We have extended Monte Carlo models based on the Dual Parton Model for
high energy hadron-nucleus collisions to the calculation of cross
sections for residual nuclei production and to nuclear
evaporation, Fermi Break-up, and high energy fission.

As it has been demonstrated in a number of past
studies~\cite{Aurenche92a,Bopp94a,Moehring91,Ranft94c}
the models used agree quite well with momentum distributions
and multiplicities of hadrons produced in high energy interactions.
Here we find in addition a quite good agreement of the average numbers
of grey prongs $\langle N_g\rangle$ and black prongs $\langle N_b\rangle$
as function of the collision
energy and as function of the target nucleus with experimental
data, which were mostly obtained in emulsion experiments.

Furthermore, calculated multiplicity distributions of grey and black
prongs agree well to data. The correlations between the fast shower
particles and grey and black prongs as well as the correlations
between grey and black prongs are often used to analyze the
observed events in terms of centrality of the collision or in
terms if the impact parameter. Our Monte Carlo events show all
of these correlations in good agreement with experimental
results.

We find a reasonable agreement of the calculated cross-sections
with data for the production of residual nuclei in most of
the mass-region below the mass of the original target nucleus.
Since our model is formulated only in terms of average nuclear properties
we can not reproduce all the fluctuations, which are
due to particular properties of individual nuclei.

At high energies we find the average numbers of grey and black
prongs to become independent from the collision energy.
This is a behaviour which can be traced back to the limiting
fragmentation property of hadron-hadron collisions in the
target or projectile rest frame. The threshold
region, where this high energy behaviour is reached is difficult
to predict in a model like ours.
As the model has enough freedom to adjust the threshold behaviour to the
behaviour of the data, a further tuning of the model parameters
might be possible as soon as more consistent data become available.
%The model has enough freedom to
%adjust the threshold behaviour to the behaviour of the data, but
%at present there are not enough data in this energy region
%and, in addition, the data often contradict each other. Therefore,
%we can only hope, that better data become available, which allow
%in future a reliable tuning of the model parameters to the
%threshold behaviour.
%
%-------------------- Acknowledgements ---------------------------------
%
\section*{Acknowledgements}
One of the authors (S.R.) acknowledges stimulating discussions with
F.W.\ Bopp and W.\ Heinrich.

%
%-------------------- Bibliography -------------------------------------
%
\clearpage
%\bibliographystyle{zpc}
%\bibliography{hep8}

%
%-------------------- Tables -------------------------------------------
%
\clearpage
\section*{Tables}
\noindent
\begin{table}[htb]
\caption{\label{pemumulttab}
    Multiplicities of grey ($N_g$), black ($N_b$), and
    heavy ($N_h=N_g+N_b$) particles in interactions of protons with
    emulsion nuclei are given for different momenta of the incident
    proton. The values as obtained with our model {\footnotesize DTUNUC} are
    compared to data from various
    experiments~\protect\cite{Fredriksson87}. Within our results we
    define ``grey particles'' as particles with a velocity $\beta=v/c$
    between 0.23 and 0.7 and therefore ``black particles''
    as particles with $\beta<0.23$. Within the experimental data
    an upper $\beta$-limit for grey particles of 0.7 is usually
    assumed,
    whereas the $\beta$-cut between ``grey'' and ``black'' may
    slightly differ between different experiments.}
\medskip
\begin{center}
\renewcommand{\arraystretch}{1.5}
\begin{tabular}{|r||c|c|c|c|c|c|} \hline
$p_{\mbox{\scriptsize Lab}}$&
\multicolumn{2}{|c|}{$\langle N_g \rangle$}&
\multicolumn{2}{|c|}{$\langle N_b \rangle$}&
\multicolumn{2}{|c|}{$\langle N_h \rangle$} \\
(GeV/$c$)& DTUNUC& Exp.& DTUNUC& Exp.& DTUNUC& Exp.
\\ \hline \hline
   6.2& 2.0& 3.58$\pm$0.11& 3.2&              & 5.2& 9.25$\pm$0.18
\\ \hline
   9.0& 2.3& 3.1 $\pm$0.4 & 3.6& 4.7 $\pm$0.5 & 5.9&
\\ \hline
  14.9& 2.4&              & 4.2&              & 6.6& 8.4
\\ \hline
  21.0& 2.6& 2.9 $\pm$0.2 & 4.6& 4.6 $\pm$0.2 & 7.2&
\\ \hline
  24.0& 2.5& 3.17$\pm$0.1 & 4.5&              & 7.0& 7.7 $\pm$0.2
\\ \hline
  50.0& 2.8& 3.07$\pm$0.1 & 5.0&              & 7.8& 7.5 $\pm$0.2
\\ \hline
  67.0& 2.9& 2.5 $\pm$0.1 & 5.2& 4.7 $\pm$0.2 & 8.1&
\\ \hline
  67.0&    & 2.85$\pm$0.09&    &              &    & 7.5 $\pm$0.2
\\ \hline
 200.0& 2.9& 2.48$\pm$0.08& 5.4& 4.79$\pm$0.12& 8.3&
\\ \hline
 300.0& 2.9&  2.6$\pm$0.2 & 5.4& 5.4          & 8.3& 7.1 $\pm$0.2
\\ \hline
 400.0& 2.9&              & 5.4&              & 8.3& 8.1 $\pm$0.2
\\ \hline
\end{tabular}
\end{center}
\end{table}
\begin{table}[htb]
\caption{\label{pcnomulttab}
    Grey ($N_g$), black ($N_b$), and heavy ($N_h=N_g+N_b$) particle
    multiplicities in interactions of protons with
    light emulsion nuclei (C,N,O) are given for different momenta
    of the incident proton. Results of the model are compared to data
    from various experiments~\protect\cite{Fredriksson87}. For the
    definition of grey and black particles we refer to the
    caption of Tab.~\protect\ref{pemumulttab}.}
\medskip
\begin{center}
\renewcommand{\arraystretch}{1.5}
\begin{tabular}{|r||c|c|c|c|c|c|} \hline
$p_{\mbox{\scriptsize Lab}}$&
\multicolumn{2}{|c|}{$\langle N_g \rangle$}&
\multicolumn{2}{|c|}{$\langle N_b \rangle$}&
\multicolumn{2}{|c|}{$\langle N_h \rangle$} \\
(GeV/$c$)& DTUNUC& Exp.& DTUNUC& Exp.& DTUNUC& Exp.
\\ \hline \hline
   6.0& 0.9& 2.54$\pm$0.06& 3.2&              & 4.1& 8.05$\pm$0.1
\\ \hline
   6.0&    & 0.96$\pm$0.07&    &              &    & 2.67$\pm$0.14
\\ \hline
   9.0& 0.86& 1.4$\pm$0.1  & 3.1&             & 3.96& 4.7$\pm$0.1
\\ \hline
  21.0& 1.02& 0.7$\pm$0.1  & 2.9& 2.2$\pm$0.1 & 3.92& 2.9$\pm$0.1
\\ \hline
  26.0& 1.05& 0.91$\pm$0.04& 2.9&             & 3.95& 2.5$\pm$0.1
\\ \hline
  50.0& 1.03& 0.91$\pm$0.04& 2.8&             & 3.83& 2.4$\pm$0.1
\\ \hline
  60.0& 1.07& 0.63$\pm$0.07& 2.9& 2.0$\pm$0.2 & 3.97& 2.6$\pm$0.2
\\ \hline
  67.0&     & 0.6$\pm$0.1  &    & 1.5$\pm$0.1 &     &
\\ \hline
  69.0& 1.05& 0.84$\pm$0.04& 2.8&             & 3.85& 3.47$\pm$0.15
\\ \hline
  69.0&     & 0.91$\pm$0.05&    &             &     & 3.65$\pm$0.1
\\ \hline
 200.0& 1.06& 0.9$\pm$0.05 & 2.8& 1.8$\pm$0.08& 3.86& 2.7$\pm$0.11
\\ \hline
 200.0&     &              &    &             &     & 2.61$\pm$0.08
\\ \hline
 200.0&     &              &    &             &     & 2.75$\pm$0.1
\\ \hline
 300.0& 1.08&              & 2.8&             & 3.88& 2.9$\pm$0.21
\\ \hline
 400.0& 1.04& 0.67$\pm$0.04& 2.8&             & 3.84& 2.47$\pm$0.09
\\ \hline
\end{tabular}
\end{center}
\end{table}
\begin{table}[htb]
\caption{\label{pagbrmulttab}
    Grey ($N_g$), black ($N_b$), and heavy ($N_h=N_g+N_b$) particle
    multiplicities in interactions of protons with
    heavy emulsion nuclei (Ag,Br) are given for different momenta
    of the incident proton. Results of the model are compared to data
    from various experiments~\protect\cite{Fredriksson87}. For the
    definition of grey and black particles we refer to the
    caption of Tab.~\protect\ref{pemumulttab}.}
\medskip
\begin{center}
\renewcommand{\arraystretch}{1.5}
\begin{tabular}{|r||c|c|c|c|c|c|} \hline
$p_{\mbox{\scriptsize Lab}}$&
\multicolumn{2}{|c|}{$\langle N_g \rangle$}&
\multicolumn{2}{|c|}{$\langle N_b \rangle$}&
\multicolumn{2}{|c|}{$\langle N_h \rangle$} \\
(GeV/$c$)& DTUNUC& Exp.& DTUNUC& Exp.& DTUNUC& Exp.
\\ \hline \hline
   9.0& 2.8 & 4.1$\pm$0.5 & 4.0 &             &6.8 &10.2$\pm$0.8
\\ \hline
  13.8& 2.9 & 6.6$\pm$0.6 & 4.7 &             &7.6 &16.0$\pm$1.4
\\ \hline
  21.0& 3.1 & 3.9$\pm$0.2 & 5.3 & 5.9$\pm$0.3 &8.4 &
\\ \hline
      &     & 3.9$\pm$0.2 &     &             &    & 9.8$\pm$0.3
\\ \hline
  24.0& 3.2 & 5.1$\pm$0.6 & 5.4 &             &8.6 &16.0$\pm$1.5
\\ \hline
  24.0&     &3.96$\pm$0.13&     &             &    & 9.5$\pm$0.3
\\ \hline
  26.0& 3.2 & 3.3$\pm$0.1 & 5.4 &             &8.6 &11.2$\pm$0.15
\\ \hline
  50.0& 3.5 &3.86$\pm$0.13& 6.0 &             &9.5 & 9.4$\pm$0.3
\\ \hline
  60.0& 3.6 & 3.4$\pm$0.2 & 6.3 & 4.9$\pm$0.6 &9.7 & 8.3$\pm$0.6
\\ \hline
  67.0& 3.5 & 3.4$\pm$0.2 & 6.1 & 6.2$\pm$0.3 &9.5 &
\\ \hline
  67.0&     & 3.1$\pm$0.1 &     &             &    & 9.7$\pm$0.3
\\ \hline
 200.0& 3.8 &3.29$\pm$0.1 & 6.6 &6.36$\pm$0.16&10.4&9.66$\pm$0.24
\\ \hline
 200.0&     &             &     &             &    &9.92$\pm$0.17
\\ \hline
 300.0& 3.9 &             & 6.8 &             &10.7& 9.9$\pm$0.5
\\ \hline
 400.0& 3.9 &             & 6.9 &             &10.8&12.4$\pm$0.9
\\ \hline
 400.0&     &3.8$\pm$0.1  &     &             &    & 9.9$\pm$0.2
\\ \hline
\end{tabular}
\end{center}
\end{table}
\begin{table}[htb]
\caption{\label{hAmulttab}
    Average grey particle multiplicity for proton, pion, and kaon
    interactions with nuclei at different energies. The data are from
    \protect\cite{Braune82}. Grey particles are defined as charged
    particles with a velocity $v=\beta c$ between $0.3c$ and $0.7c$.}
\medskip
\begin{center}
\renewcommand{\arraystretch}{1.5}
\begin{tabular}{|c||c|c|c|c|c|c|} \hline
&
\multicolumn{2}{|c|}{$E_{\mbox{\scriptsize Lab}}$=50~GeV}&
\multicolumn{2}{|c|}{$E_{\mbox{\scriptsize Lab}}$=100~GeV}&
\multicolumn{2}{|c|}{$E_{\mbox{\scriptsize Lab}}$=150~GeV} \\
& DTUNUC& Exp.& DTUNUC& Exp.& DTUNUC& Exp.
\\ \hline \hline
 p--C & 0.74& 0.91$\pm$0.05& 0.73& 0.82$\pm$0.04& 0.77& 0.89$\pm$0.04
\\ \hline
 p--Cu& 2.2 & 2.23$\pm$0.11& 2.2 & 2.26$\pm$0.11& 2.3 & 2.27$\pm$0.11
\\ \hline
 p--Pb& 4.6 & 4.04$\pm$0.2 &     &              & 4.5 & 3.75$\pm$0.19
\\ \hline
 $\pi^+$--C & 0.69& 0.85$\pm$0.04& 0.69& 0.81$\pm$0.04& 0.64& 0.84$\pm$0.04
\\ \hline
 $\pi^+$--Cu& 1.96& 1.99$\pm$0.1 & 1.98& 2.04$\pm$0.1 & 1.97& 1.99$\pm$0.1
\\ \hline
 $\pi^+$--Pb& 4.0 & 3.42$\pm$0.17& 3.9 & 2.89$\pm$0.14& 3.9 & 3.31$\pm$0.17
\\ \hline
 $K^+$--C & 0.65& 0.81$\pm$0.04&     &              & 0.67& 0.80$\pm$0.04
\\ \hline
 $K^+$--Cu& 1.82& 1.92$\pm$0.1 &     &              & 1.76& 1.93$\pm$0.1
\\ \hline
 $K^+$--Pb& 3.62& 3.43$\pm$0.17&     &              & 3.77& 3.23$\pm$0.16
\\ \hline
\end{tabular}
\end{center}
\end{table}
\begin{table}[htb]
\caption{\label{pAmulttab}
    Dependence of the average multiplicities of grey ($N_g$) and
    heavy ($N_h$) particles on the target mass number in
    proton-nucleus interactions at
    $p_{\mbox{\scriptsize Lab}}=300$~GeV/$c$. Results of the model are
    compared to data
    from various experiments~\protect\cite{Fredriksson87}. ``Grey'' and
    ``heavy'' are defined as given in the caption of
    Tab.~\protect\ref{pemumulttab}}
\medskip
\begin{center}
\renewcommand{\arraystretch}{1.5}
\begin{tabular}{|c||c|c|c|c|} \hline
Nucleus&
\multicolumn{2}{|c|}{$\langle N_g \rangle$}&
\multicolumn{2}{|c|}{$\langle N_h \rangle$} \\
& DTUNUC& Exp.& DTUNUC& Exp.
\\ \hline \hline
C,N,O    & 1.1&              &3.9 & 2.9$\pm$0.21 \\ \hline
Al       & 1.4&              &5.4 &              \\ \hline
Cr       & 2.1&              &7.3 & 7.2$\pm$0.7  \\ \hline
Emulsion & 2.6& 2.6$\pm$0.2  &7.9 & 7.1$\pm$0.2  \\ \hline
Ag,Br    & 3.2&              &9.7 & 9.9$\pm$0.5  \\ \hline
W        & 5.0&              &13.5&12.9$\pm$1.2  \\ \hline
Au       & 5.4&              &14.2&              \\ \hline
\end{tabular}
\end{center}
\end{table}
%
%-------------------- Figures ------------------------------------------
%
\clearpage
\section*{Figure Captions}
\begin{enumerate}
\item \label{pAex1}
      Average excitation energies of residual target nuclei in
      proton--nucleus interactions before evaporation
      are shown for different momenta of the incident proton (a).
      In (b) the average excitation energies are given per
      nucleon of the residual target nucleus.
\item \label{pAex2}
      Target mass dependence of the average excitation energies for
      residual target nuclei with mass $A_{\mbox{\scriptsize res}}$
      in proton--nucleus interactions before evaporation.
      $\Delta A$ is the number of nucleons lost by the target nucleus
      in the primary collision ($\Delta A=A_t-A_{\mbox{\scriptsize res}}$).
\item \label{pau300ex}
      In a) the distribution of excitation energies of gold prefragments in
      proton-gold interactions at 300~GeV/$c$ is shown for different
      ranges of prefragment mass $A_{\mbox{\scriptsize res}}$.
      The corresponding excitation energy distributions per nucleon of the
      gold prefragment are given in b).
\item \label{pemumult}
      Mean multiplicities of shower, grey, and heavy particles in
      collisions of protons with emulsion nuclei. Data from various
      experiments~\protect\cite{Fredriksson87} (points) are compared
      to results of the model (lines).
\item \label{pcnoagbrmult}
      Mean multiplicities of shower, grey, and heavy particles in
      collisions of protons with emulsion nuclei are shown for the
      component consisting of light nuclei (C,N,O) (a) and
      the heavy component (Ag, Br) (b). Data from various
      experiments~\protect\cite{Fredriksson87} (points) are compared
      to results of the model (lines).
\item \label{hemung}
      Grey particle multiplicity distributions for interactions of
      protons (a) and $\Sigma^-$--hyperons (b) with emulsion nuclei
      are plotted together with experimental
      results~\protect\cite{Stenlund82,Szarska93}.
\item \label{hemunb}
      Black particle multiplicity distributions for interactions of
      protons (a) and $\Sigma^-$--hyperons (b) with emulsion nuclei
      are plotted together with experimental
      results~\protect\cite{Stenlund82,Szarska93}.
\item \label{pemunbnscorr}
      The correlations between grey ($N_g$) and shower ($N_s$) particle
      multiplicities (a,b) and black ($N_b$) and shower particle
      multiplicities (c,d) in interactions
      of protons with emulsion nuclei are compared to experimental
      results~\protect\cite{ALMT75}.
\item \label{pemungnbcorr}
      The correlations between grey ($N_g$) and black ($N_b$) particle
      multiplicities in interactions of protons with emulsion nuclei
      are compared to experimental
      results~\protect\cite{ALMT75,Dabrowska93a}.
\item \label{pAmult}
      Target mass number dependence of the average multiplicities of
      shower, grey, and heavy particles in proton--nucleus interactions.
      Data from various experiments~\protect\cite{Fredriksson87} (points)
      are compared to results of the model (lines).
\item \label{pAngwa80}
      The distributions of grey particle multiplicities in
      proton--nucleus (a) and pion--nucleus (b) interactions as
      calculated with {\footnotesize DTUNUC} are compared to experimental
      results of the WA80-Collab.~\protect\cite{Albrecht93}.
\item \label{pagresA}
      Mass distributions of prefragments produced in proton-silver
      interactions at 11.5~GeV (a) and at 300~GeV (b) as obtained with the
      model are compared to experimental results~\cite{English74,Porile79}.
\item \label{pauresA}
      Charge distributions of prefragments produced in proton-gold
      interactions at 10.6~GeV (a) and mass distributions of
      prefragments produced in proton-gold interactions at 800~GeV (b)
      as obtained with the model are compared to experimental
      results~\cite{Heinrich95,Sihver92}. In addition, the distributions
      obtained without high-energy fission are shown (crosses).
\item \label{pAuZiso}
      The yield distributions from interactions of protons with $^{197}$Au
      nuclei at 800~GeV are shown together with experimental results
      of Sihver et al.~\cite{Sihver92} for three different intervals of
      the mass of the prefragment.
\item \label{pAuprcl}
      The total momentum of residual nuclei as a function of the mass
      loss of the target nucleus are compared to experimental
      results. The experimental data are from different experiments and
      have been taken from Fig.~7 in~\cite{Sihver92}.
\end{enumerate}

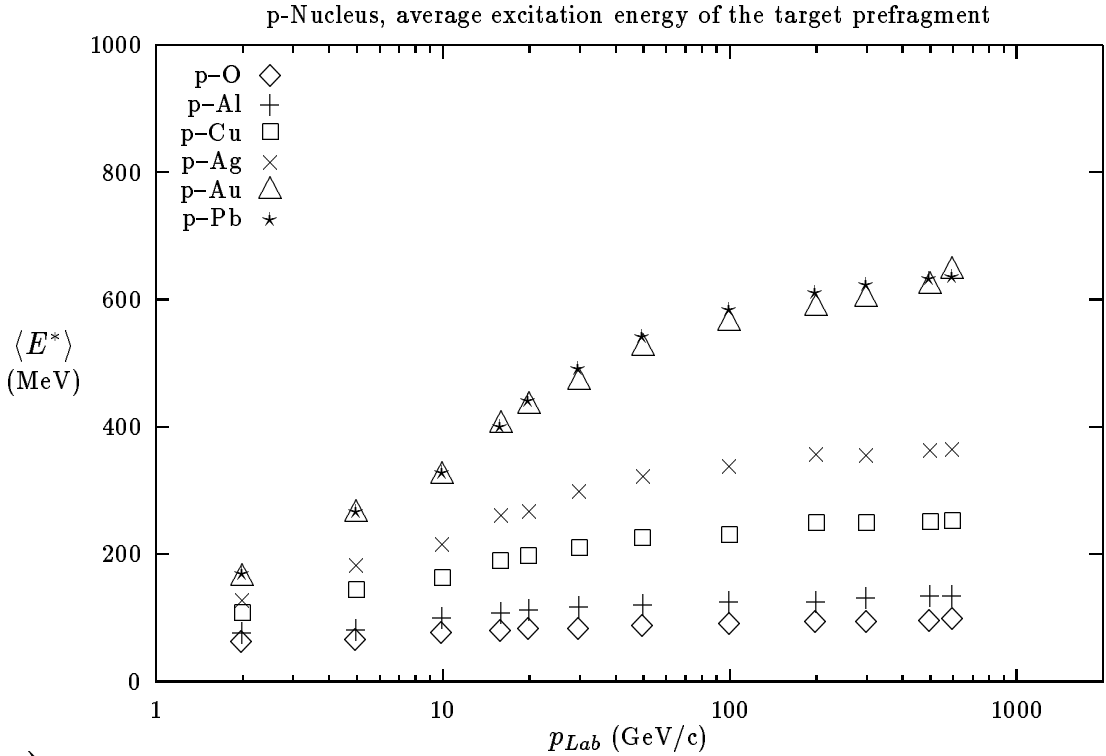
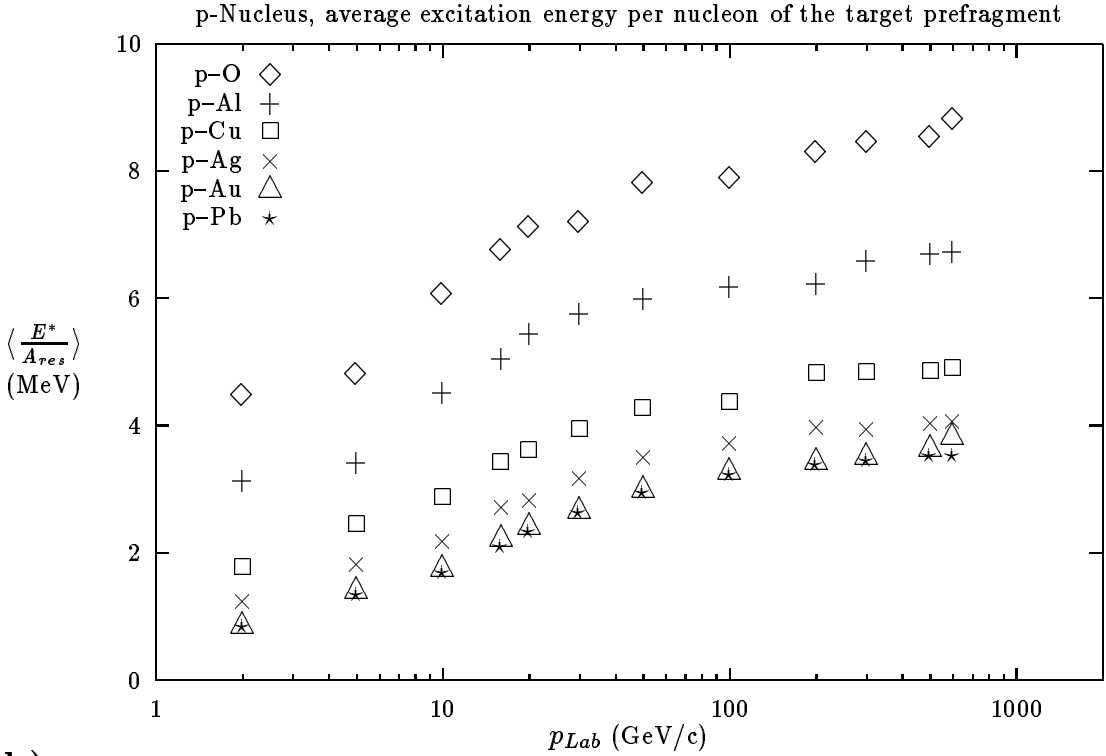
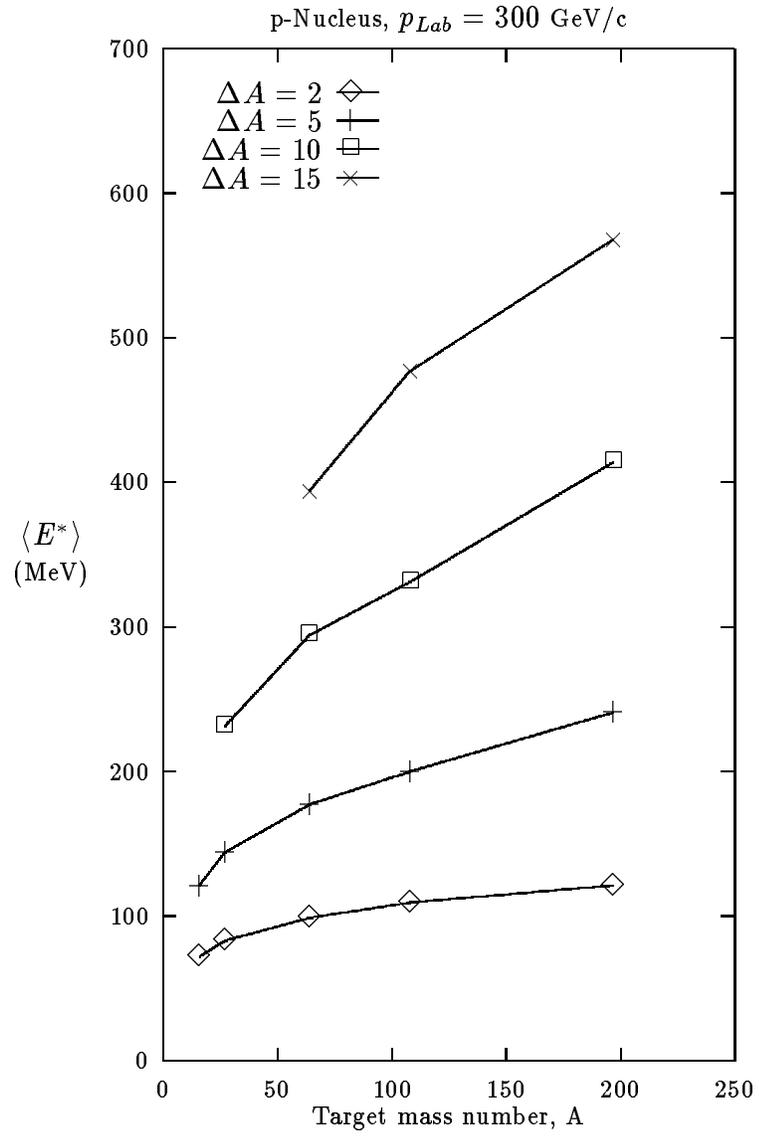
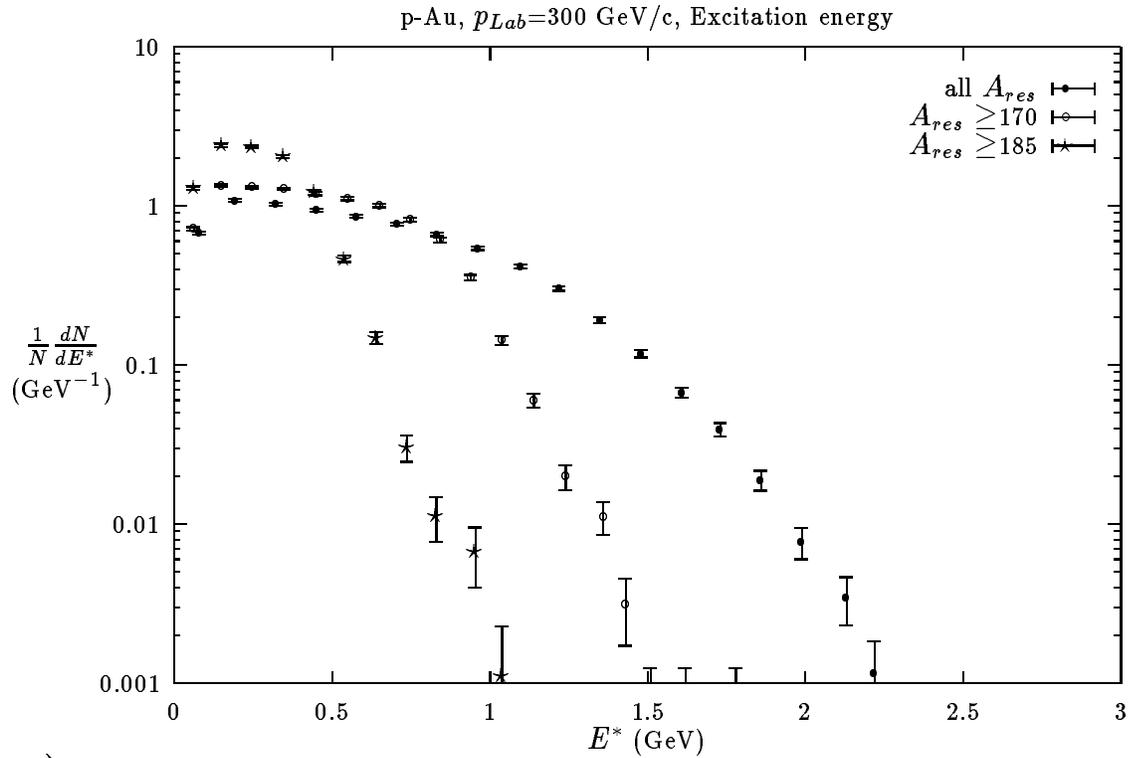
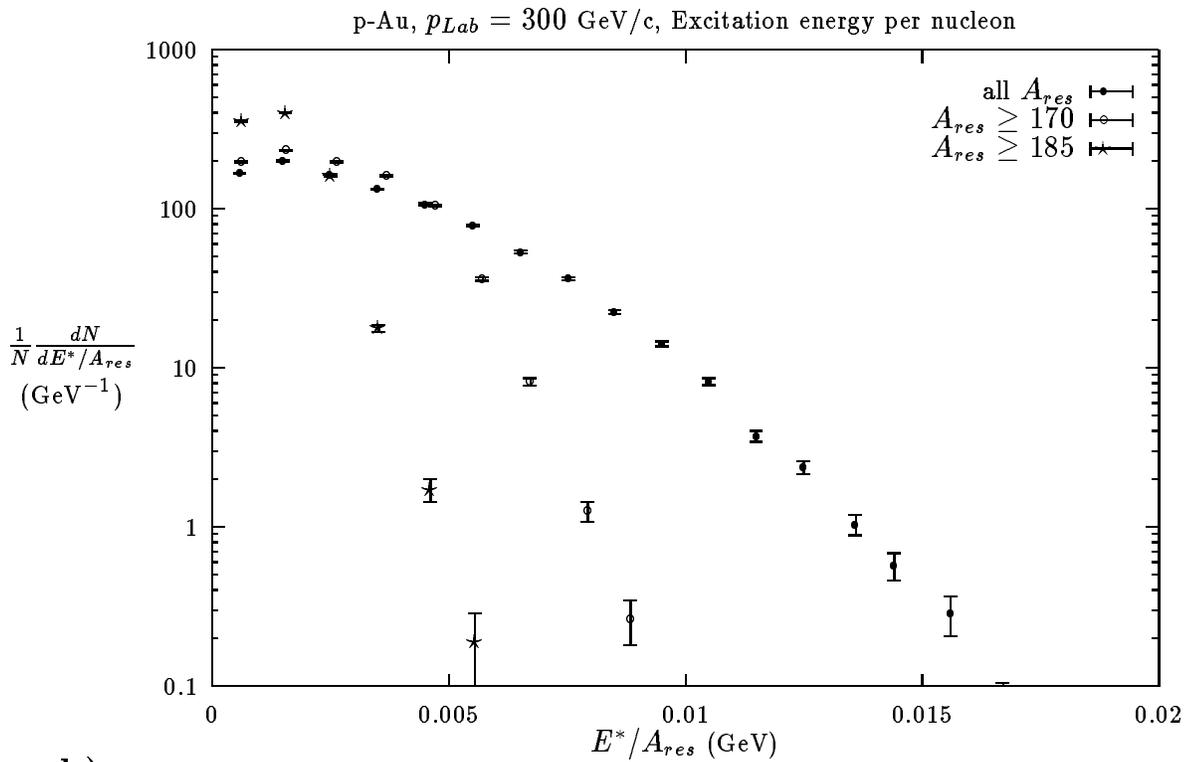
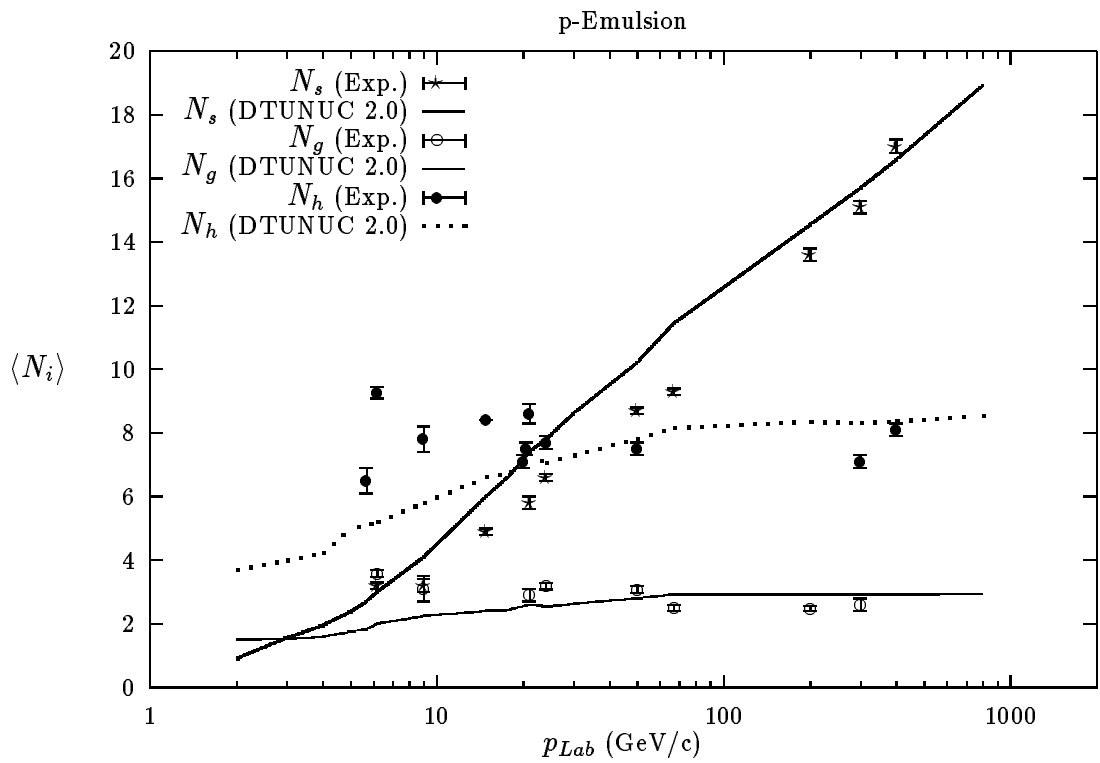
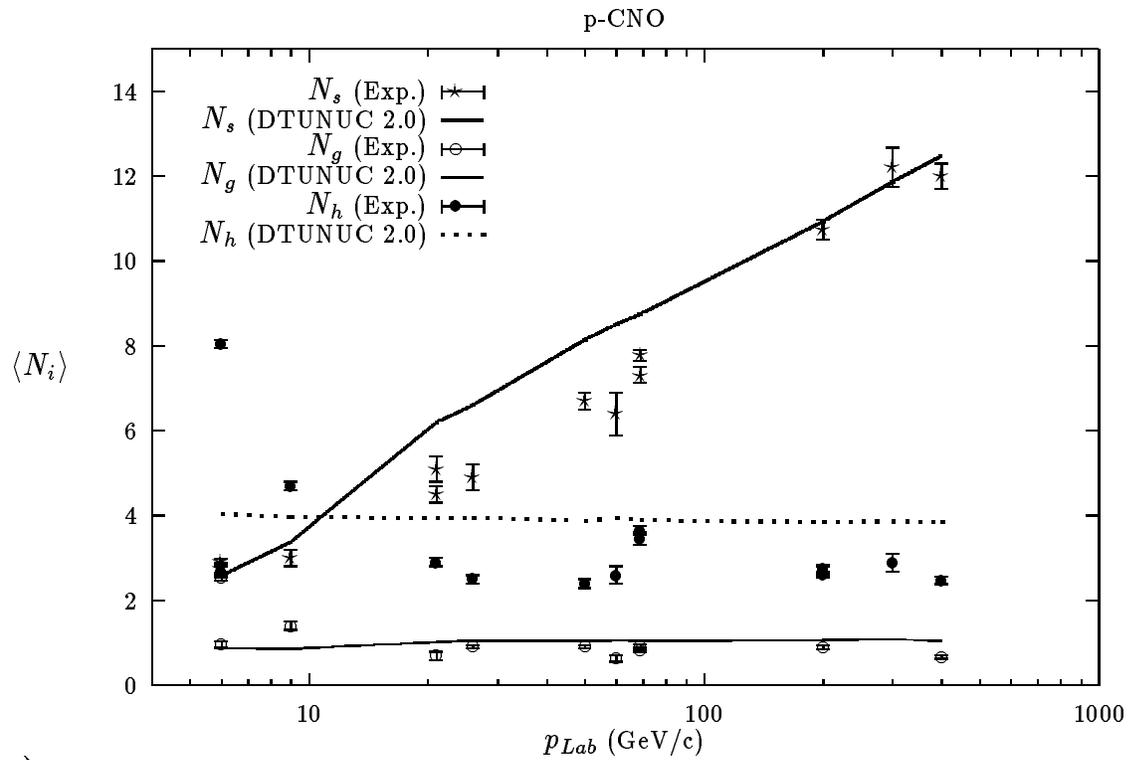
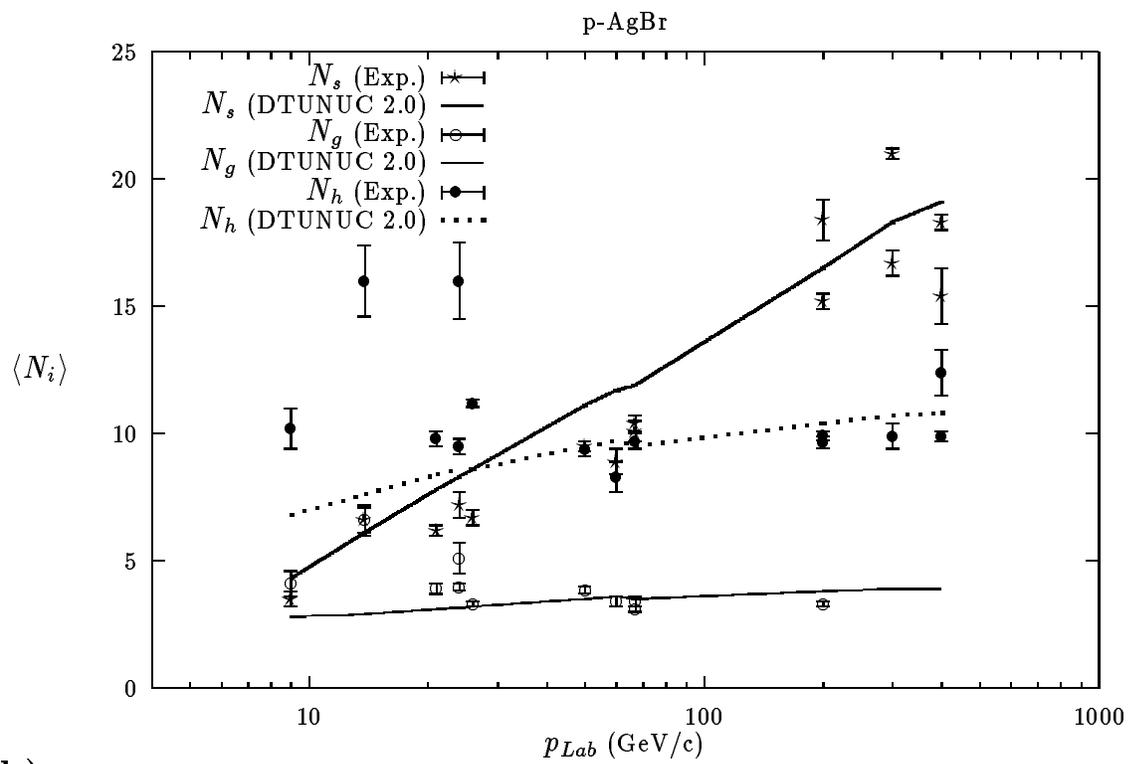
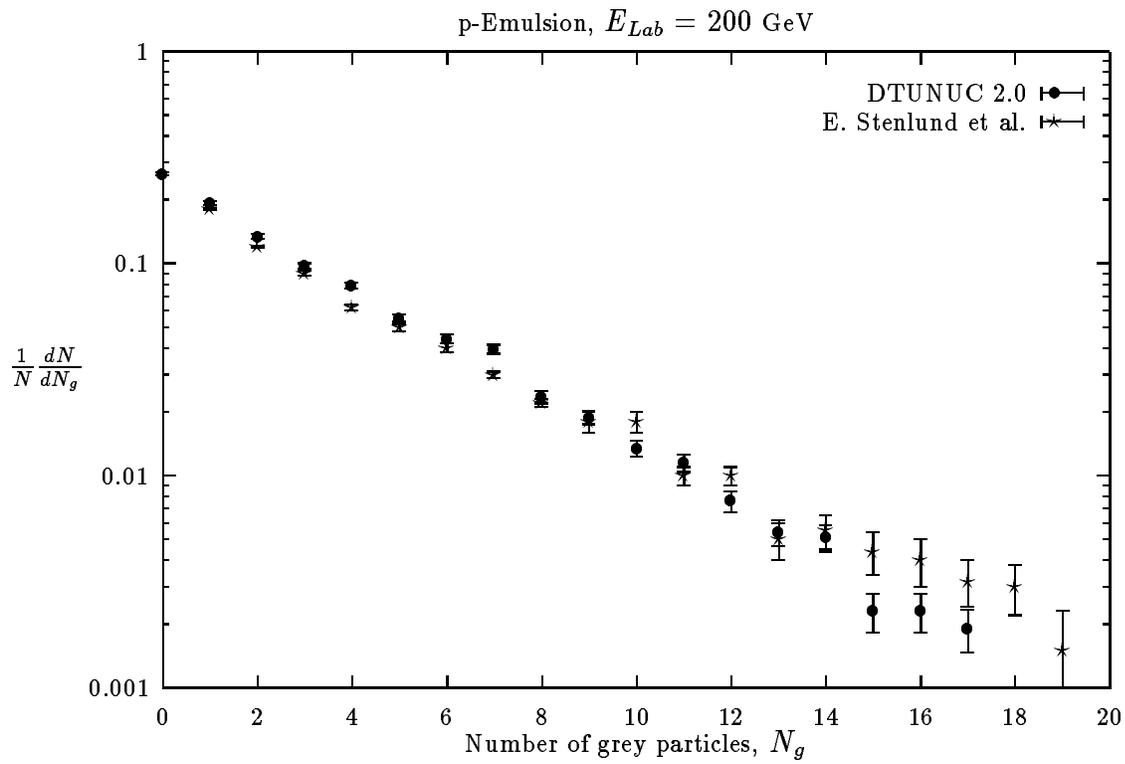
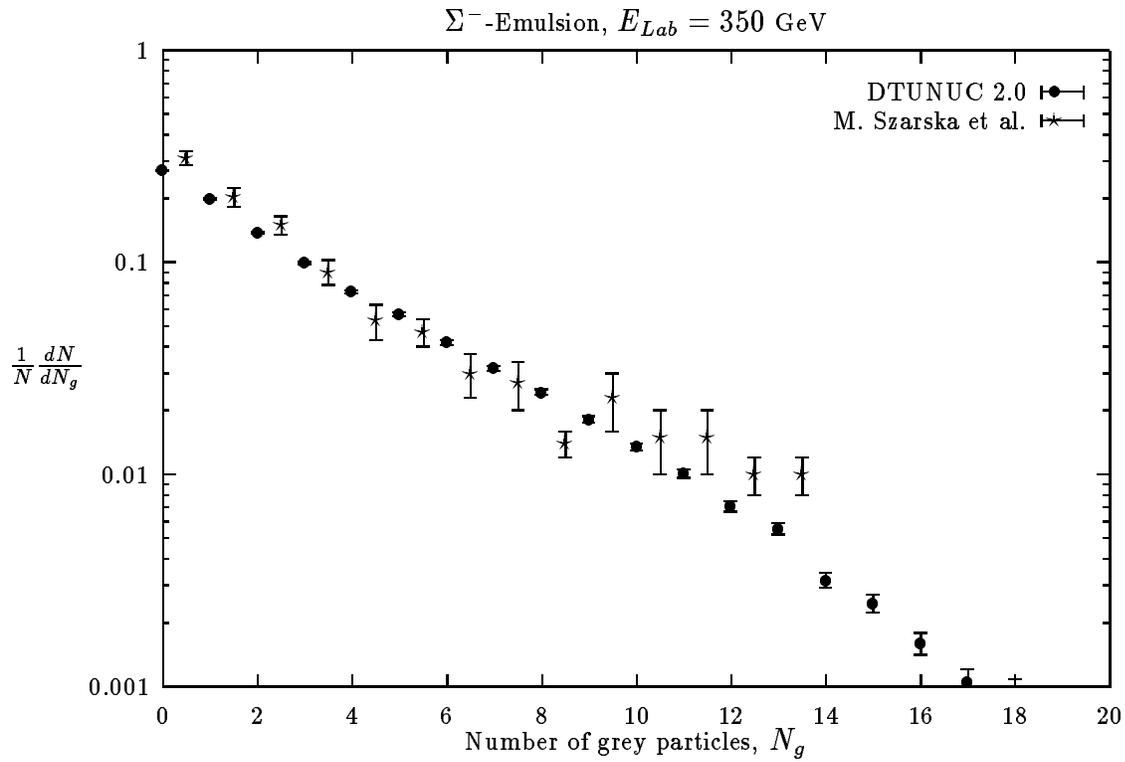
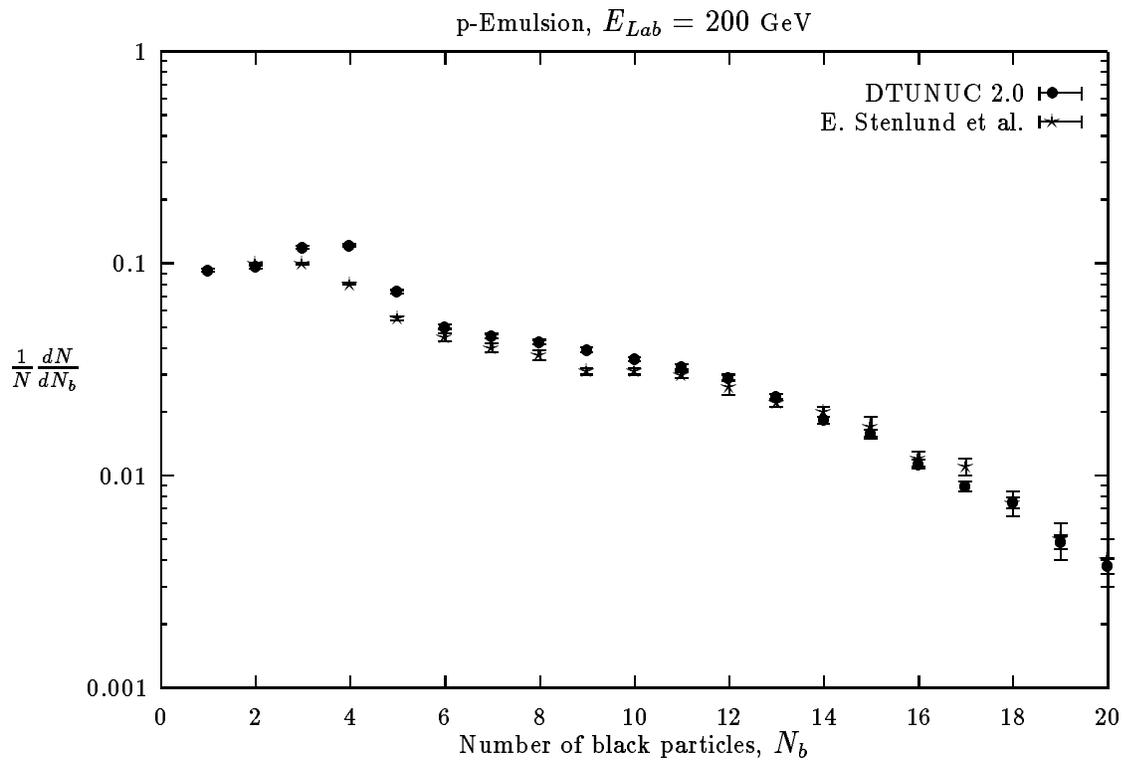
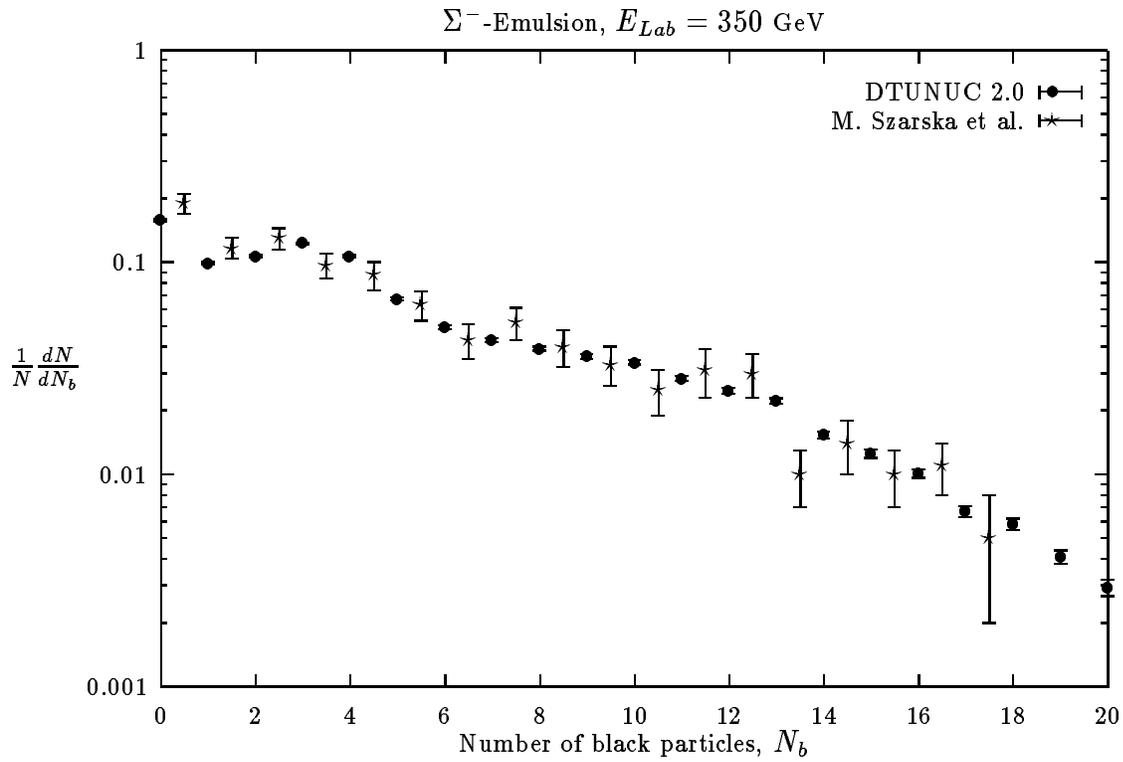
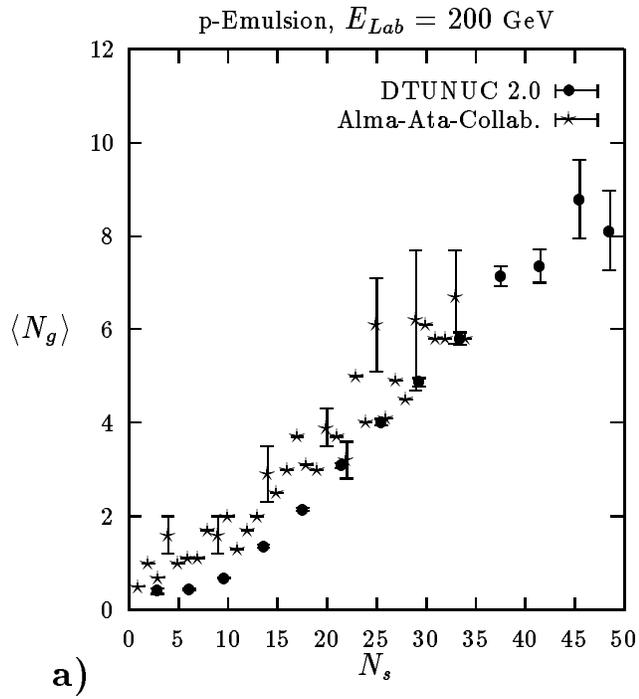
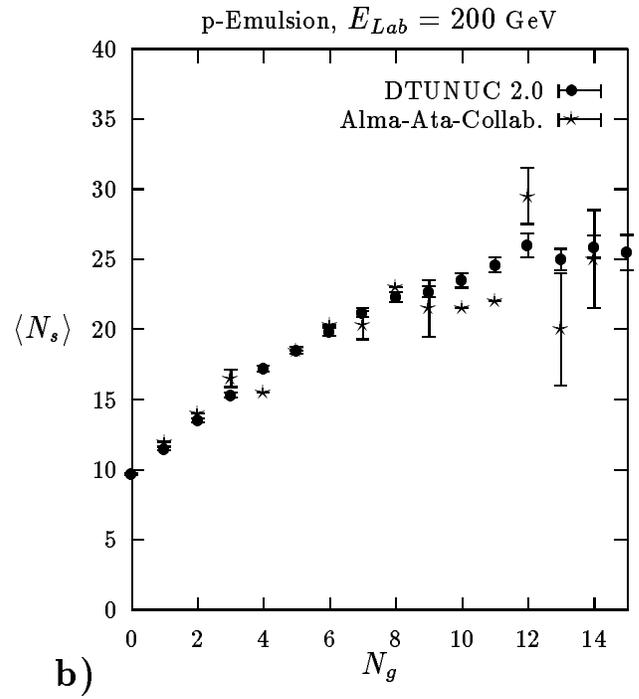
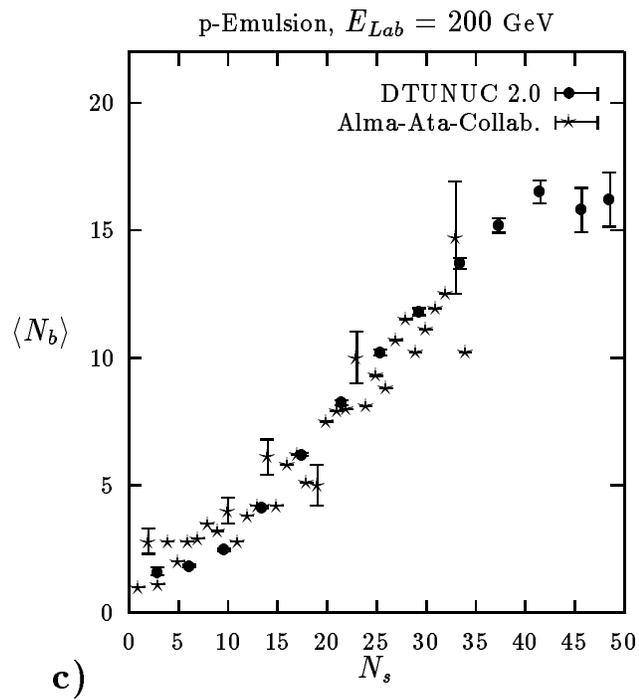
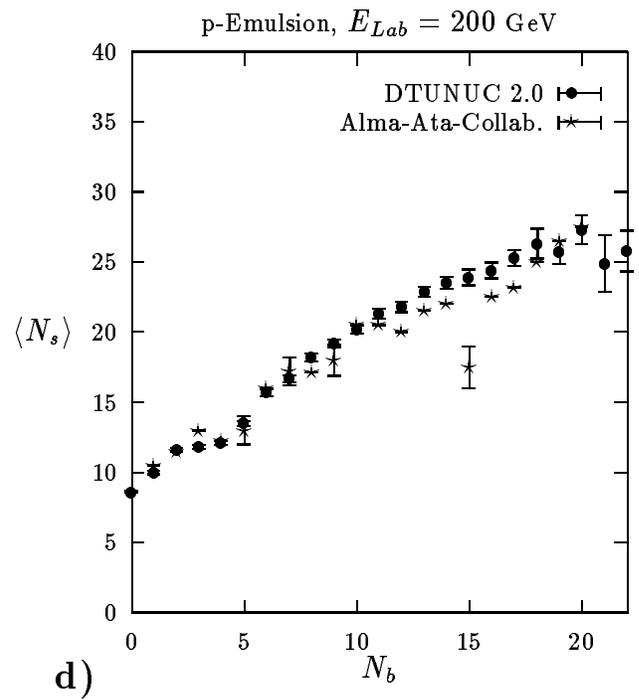
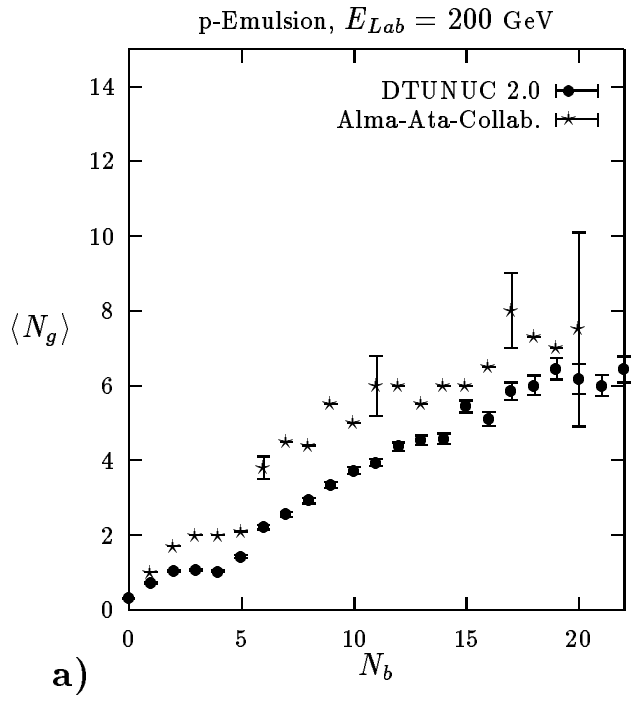
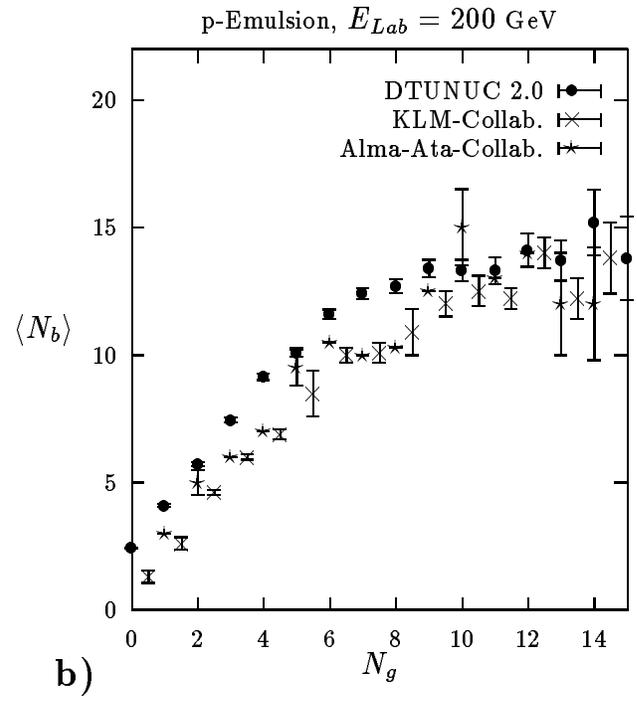
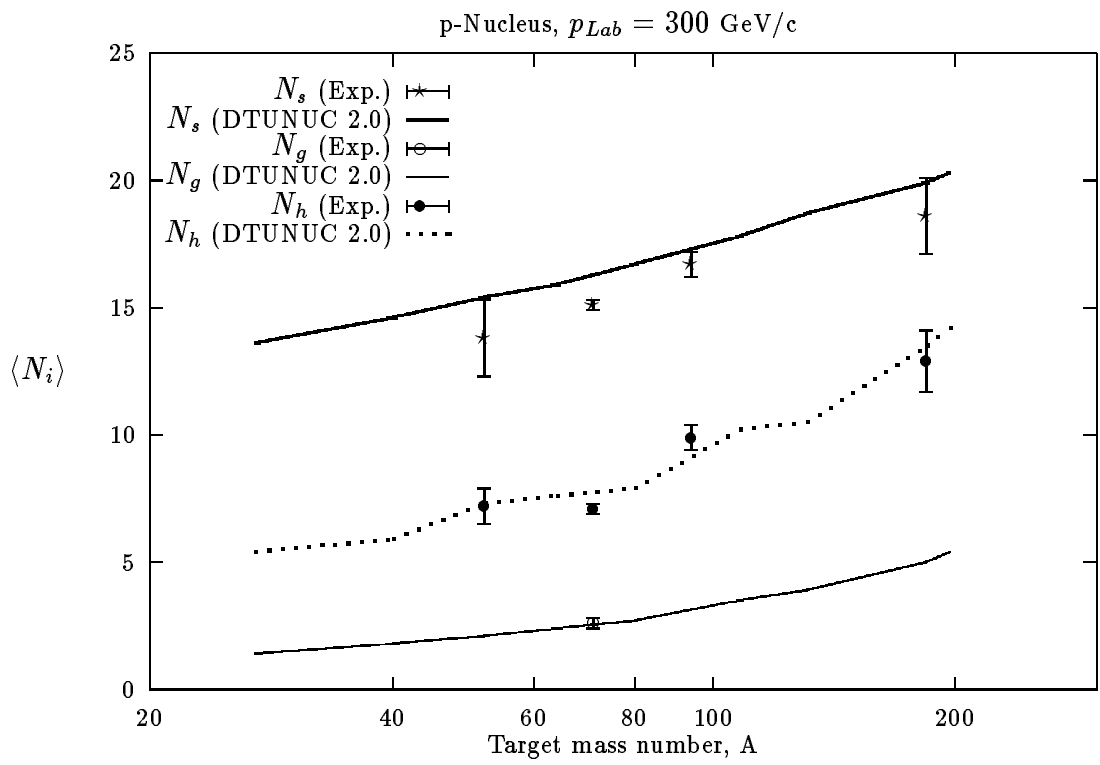
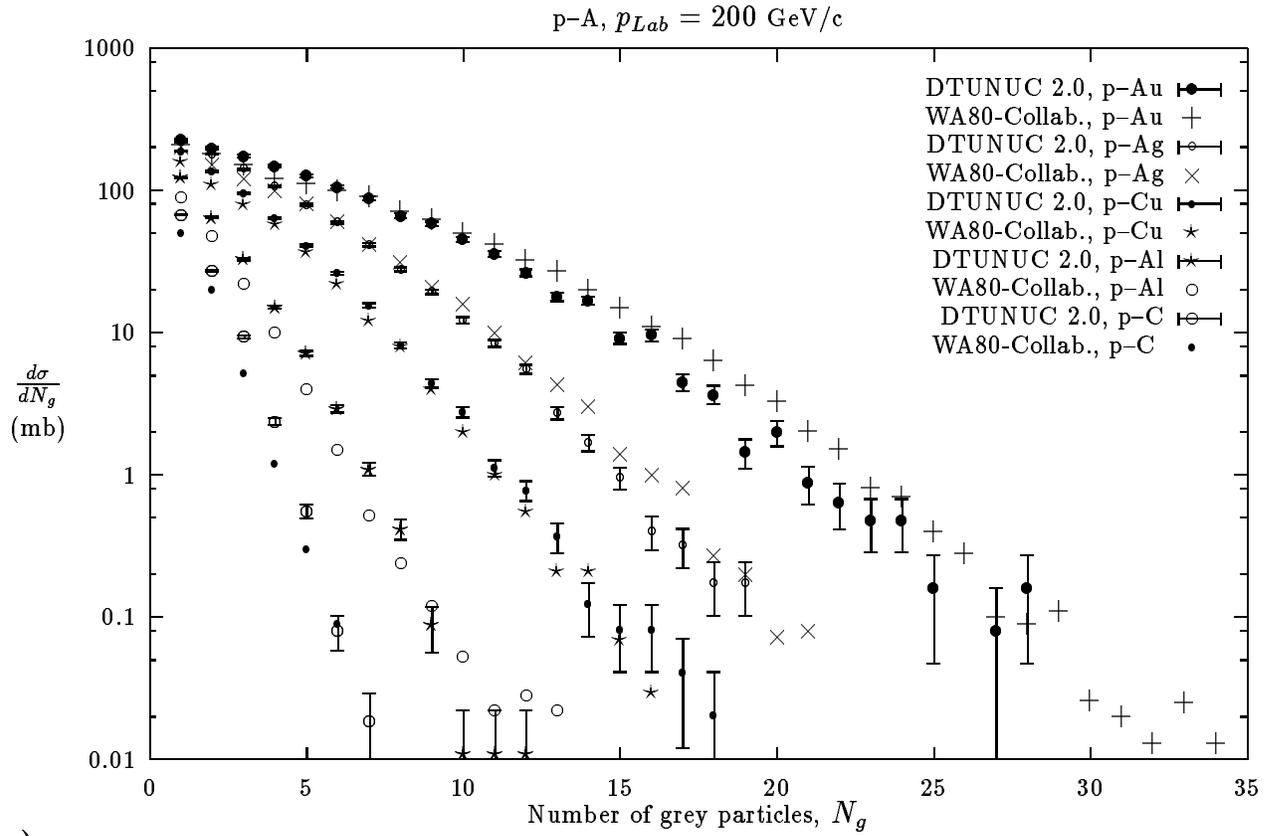
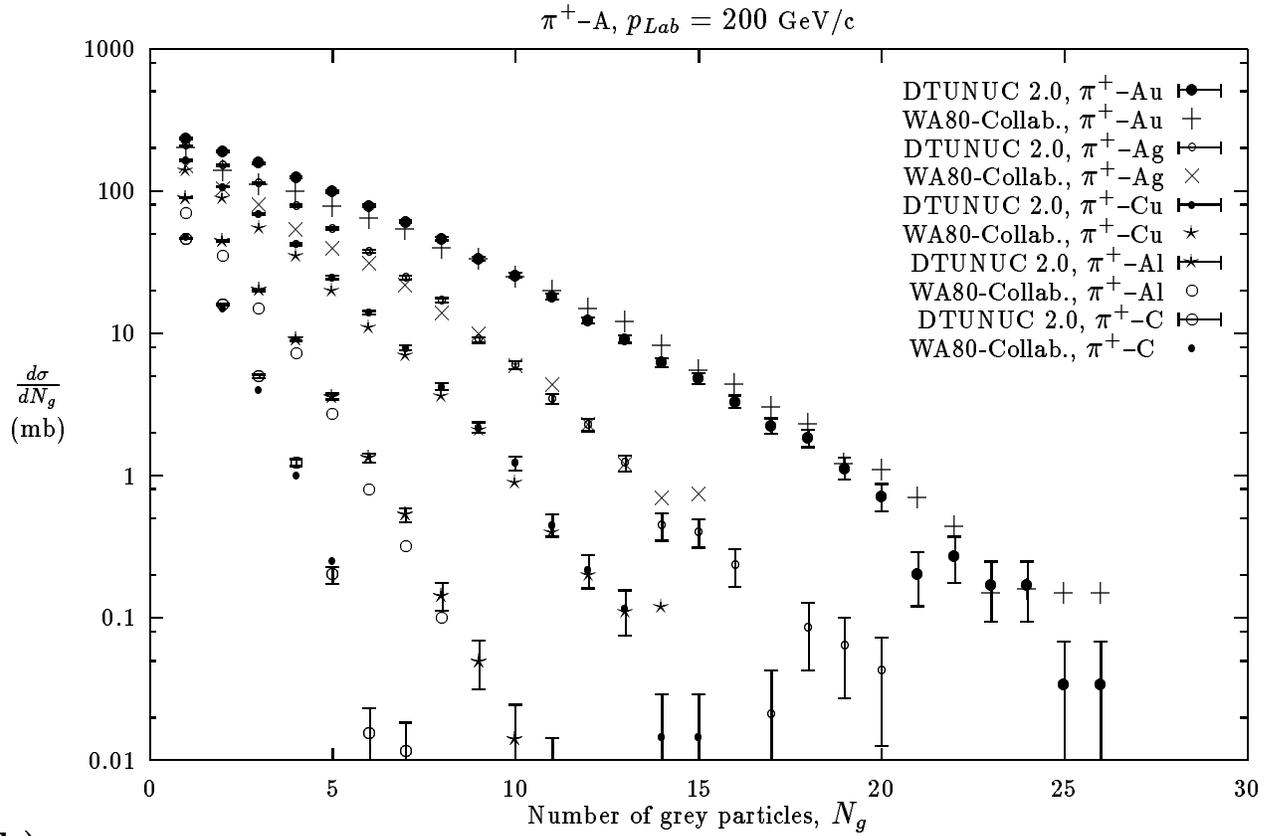
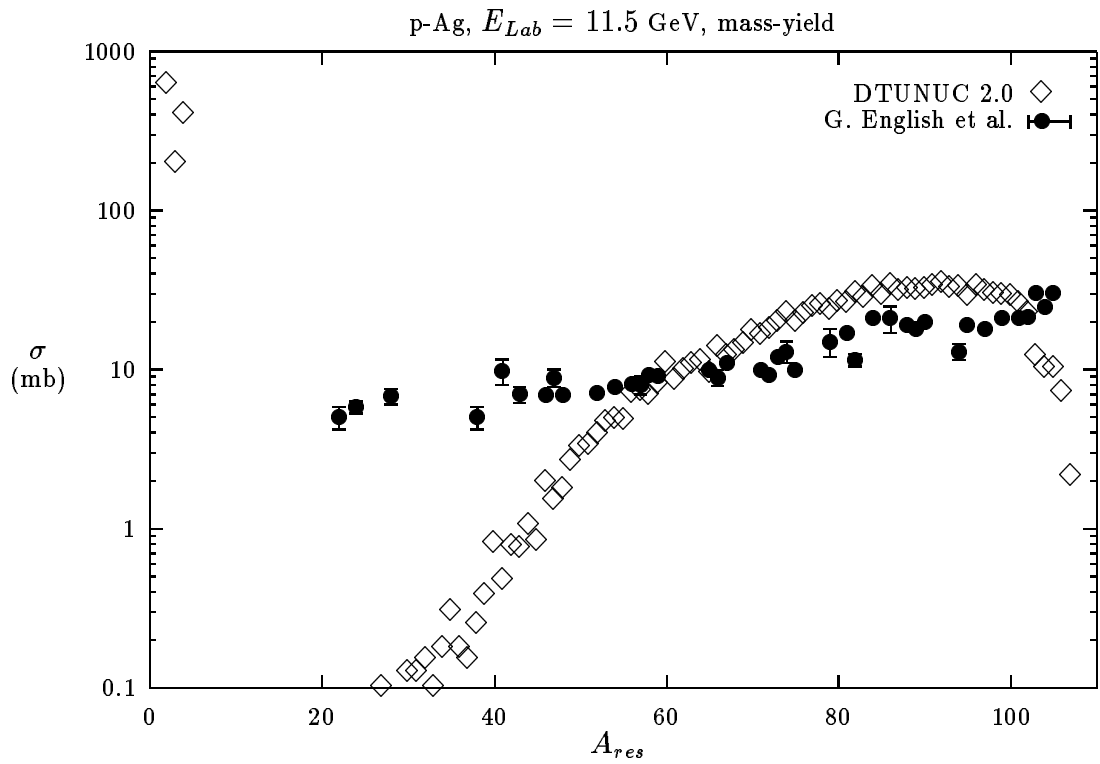
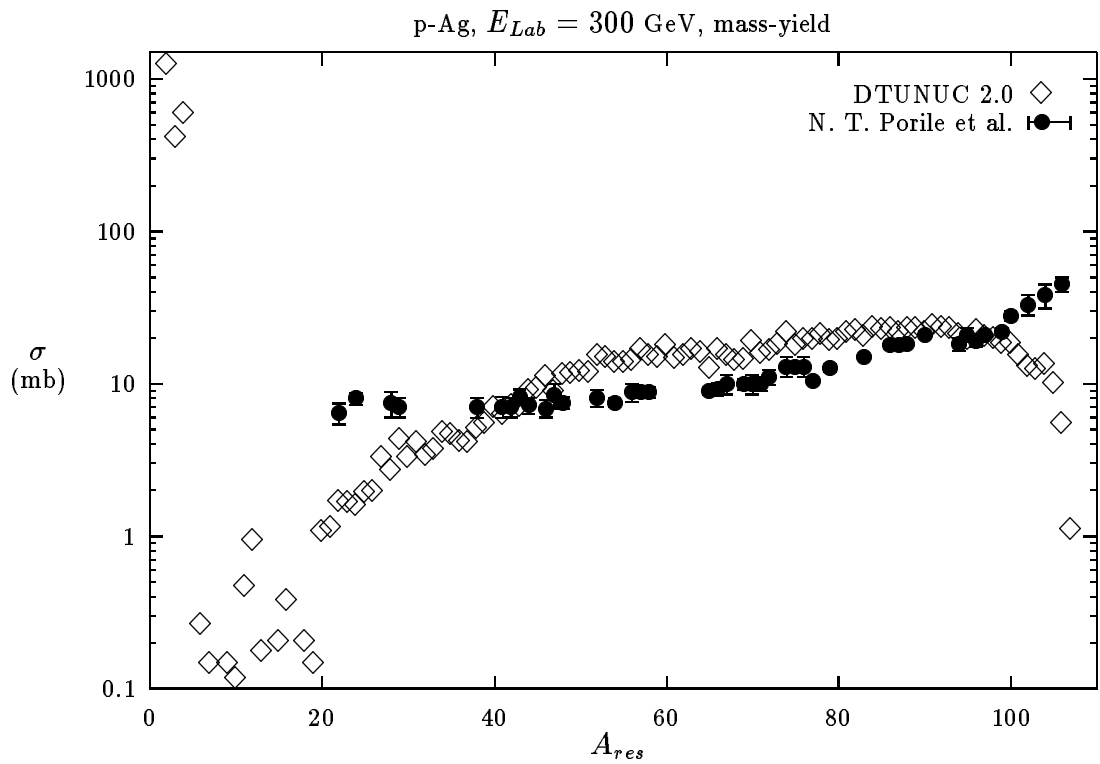
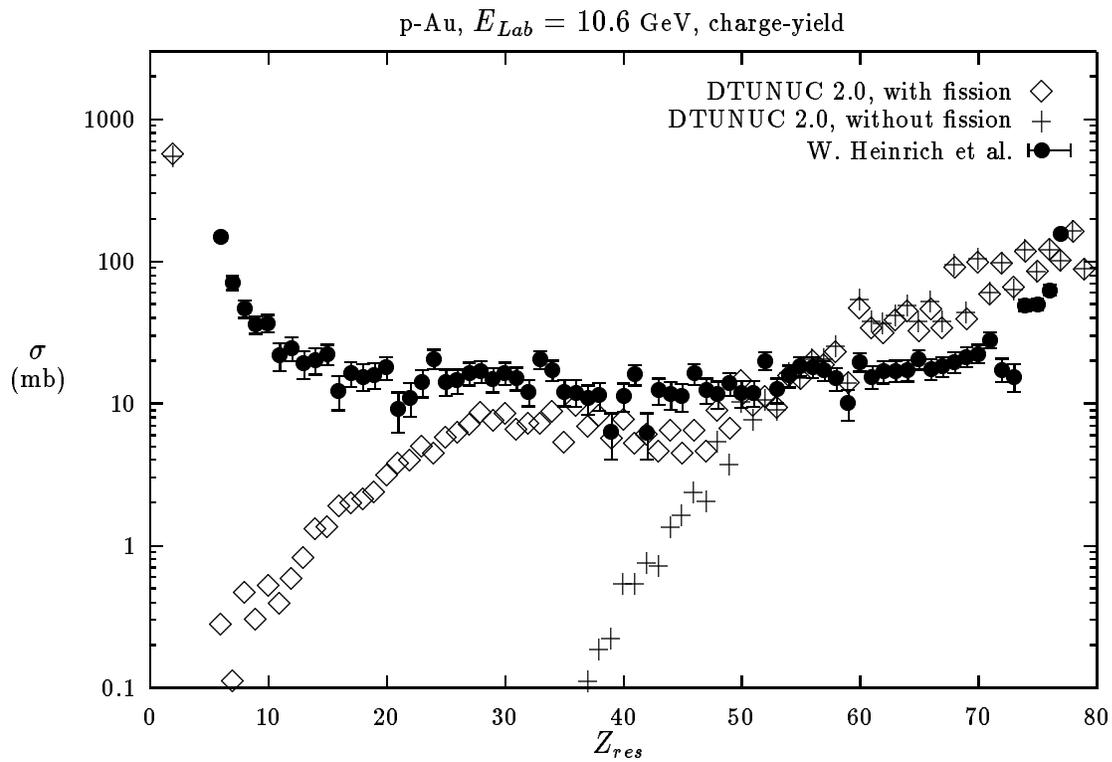
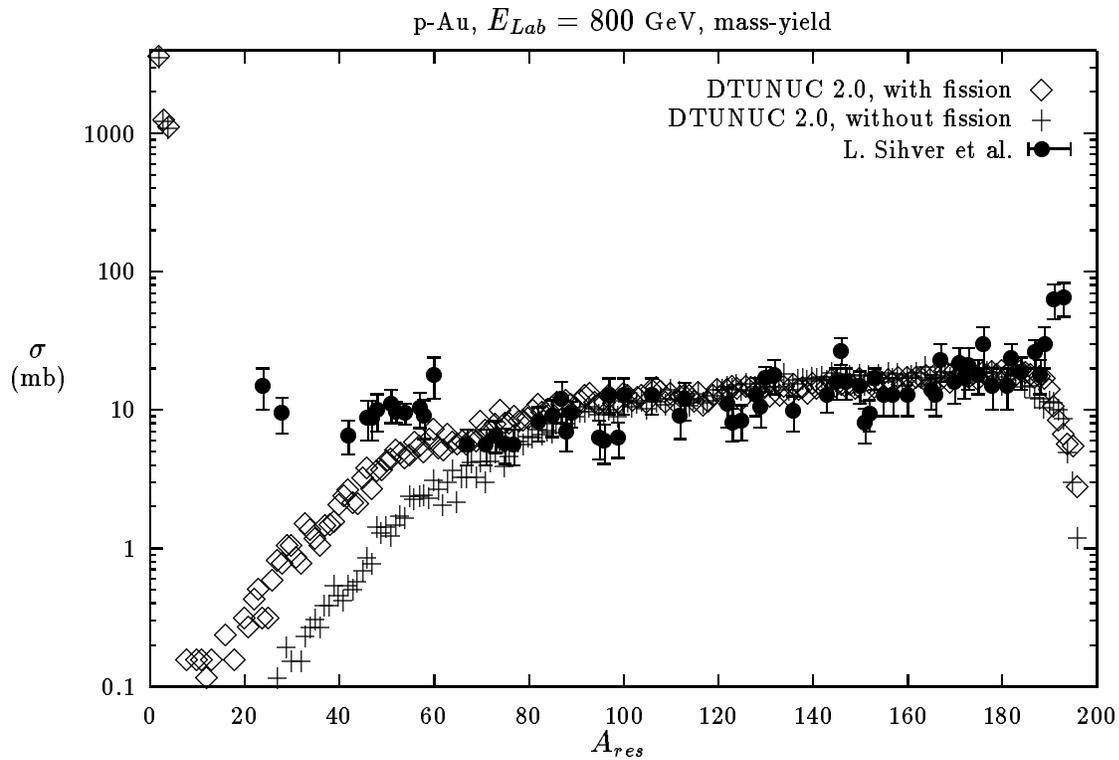
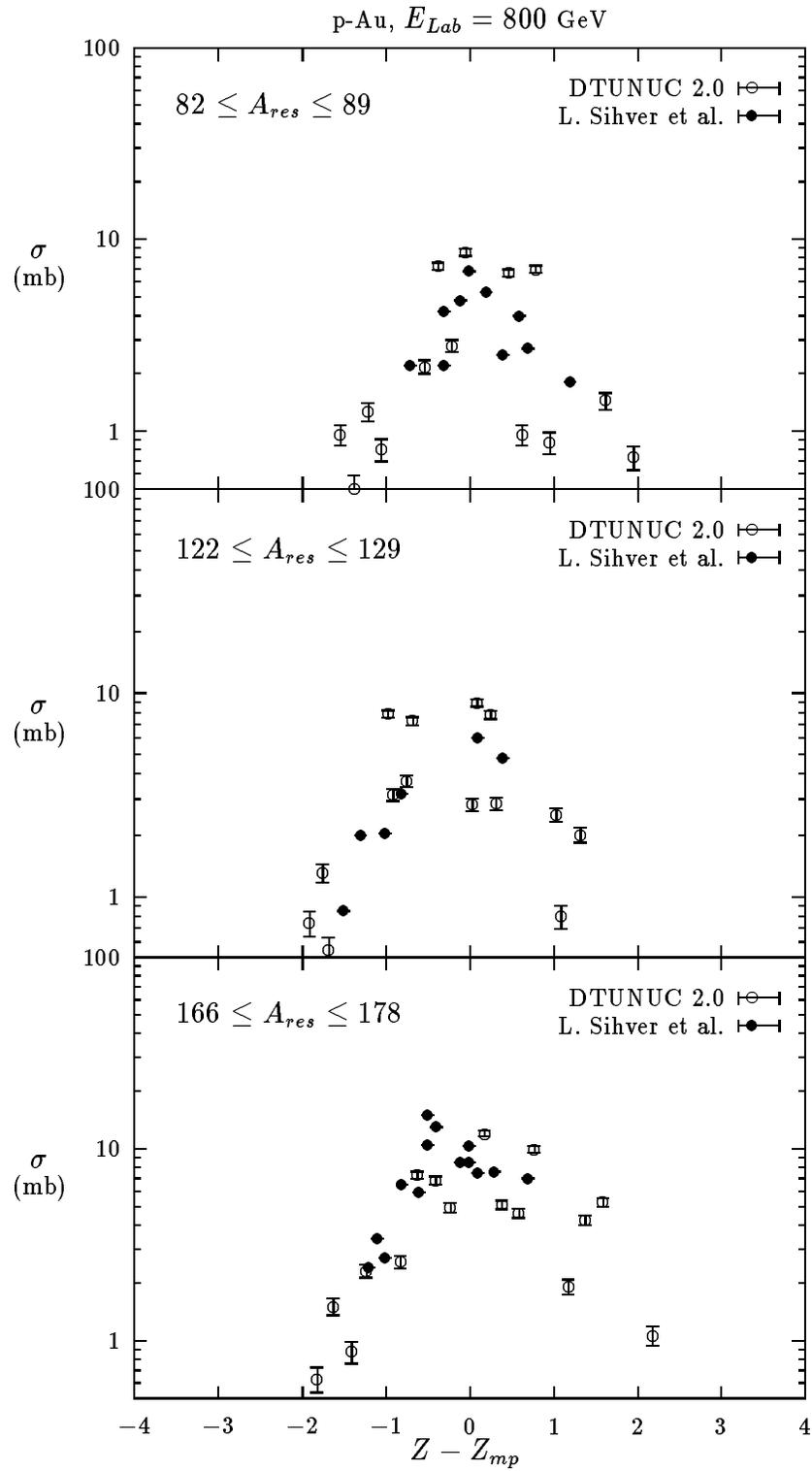
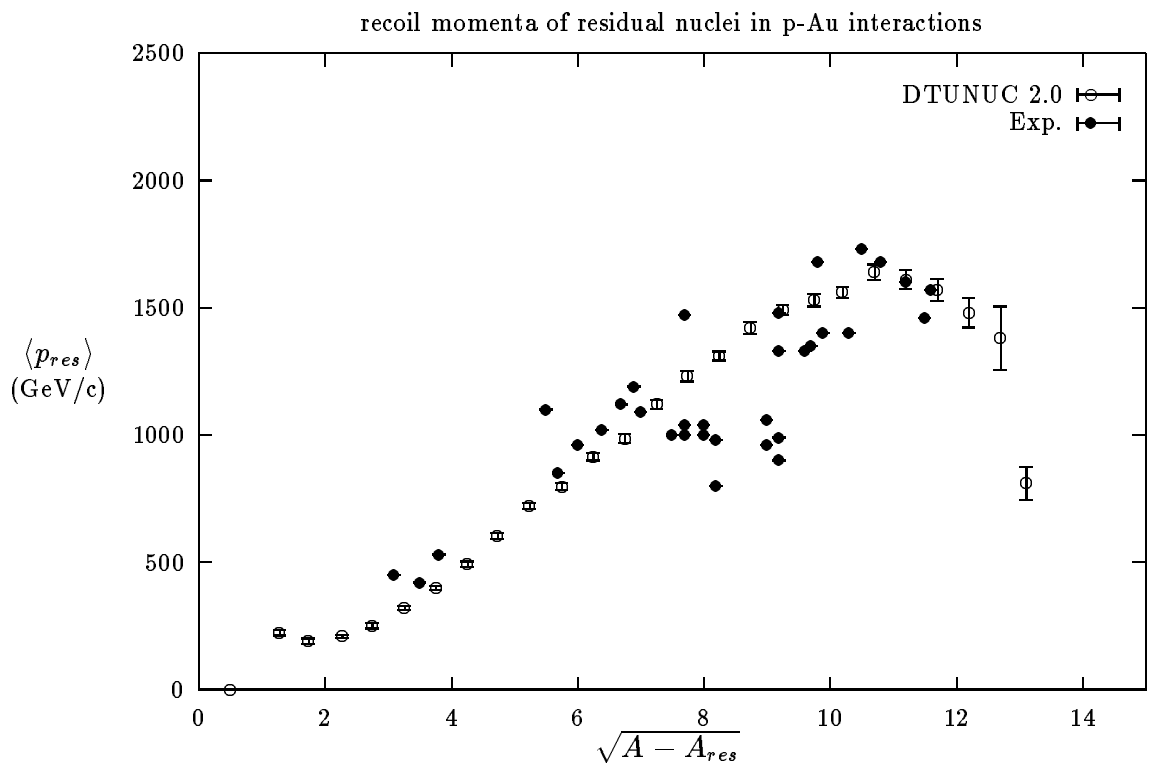
\end{document}